\documentclass[12pt,a4paper]{article}
\usepackage[utf8]{inputenc}
\usepackage{amsmath}
\usepackage{amsfonts}
\usepackage{amssymb}
\usepackage{mathtools}
\usepackage{graphicx}
\usepackage{cite}
\usepackage[capitalise]{cleveref}
\graphicspath{ {./figures/} }
\usepackage[left=2cm,right=2cm,top=2cm,bottom=2cm]{geometry}
\crefformat{section}{#2section~#1#3}

\author{Tetiana Obikhod and Ievgenii Petrenko}
\title{\bf{Studying of final states in p-Au and p-Pb collisions}}
\date{%
    {\it Institute for Nuclear Research NAS of Ukraine, Kyiv 03028, Ukraine}\\%
    \today
}
\begin{document}

\maketitle

\section{Abstract}

	In the framework of PYTHIA8.2 program we considered p-Pb and p-Au heavy ion collisions at the energy of 5.02 TeV and 8 TeV. The
advantage of this program is in the combining of several nucleon-nucleon collisions into one heavy ion collision, based on phenomenological treatment of a hadron as a vortex line in a colour superconducting medium, the consistent treatment of the central rapidity region with improvements of Glauber-like model where diffractive excitation processes are taken into account. We have considered the influence of impact parameter correlations on the production cross
sections of p-Pb and p-Au processes to estimate the influence of hard and soft subprocesses on basic hadronic final-state properties in proton-ion collisions. Using these characteristics based on semi-hard multiparton interaction model we received the transverse momentum
and rapidity distributions of K-meson and Lambda baryon at the energy of 5.02 TeV and 8.14 TeV.

\section{Introduction}

The studying for hard processes through the few-body hard-scattering processes which lead to complex multiparticle final states is one of the interesting goal of high energy physics. In spite of the standard basic processes there will remain some part of the internal processes, involving parton showers, processes with extra dimensions, supersymmetric processes and more new states. 
\par
We will concentrate on QCD hard processes of pp, p-Pb and p-Au types with higher final-state multiplicity, which include soft- and hard-QCD processes. If low-mass diffractive systems are represented as non-perturbative hadronizing strings, then for diffractive systems with masses about 10 GeV, multiparton interactions (MPI) - multiple interactions between several pairs of incoming partons presented by initial-state radiation (ISR), final-state radiation (FSR), and string fragmentation are included. The schematic representation of the mentioned processes is presented in Fig.1

\begin{center}
 \includegraphics[width=0.5\textwidth]{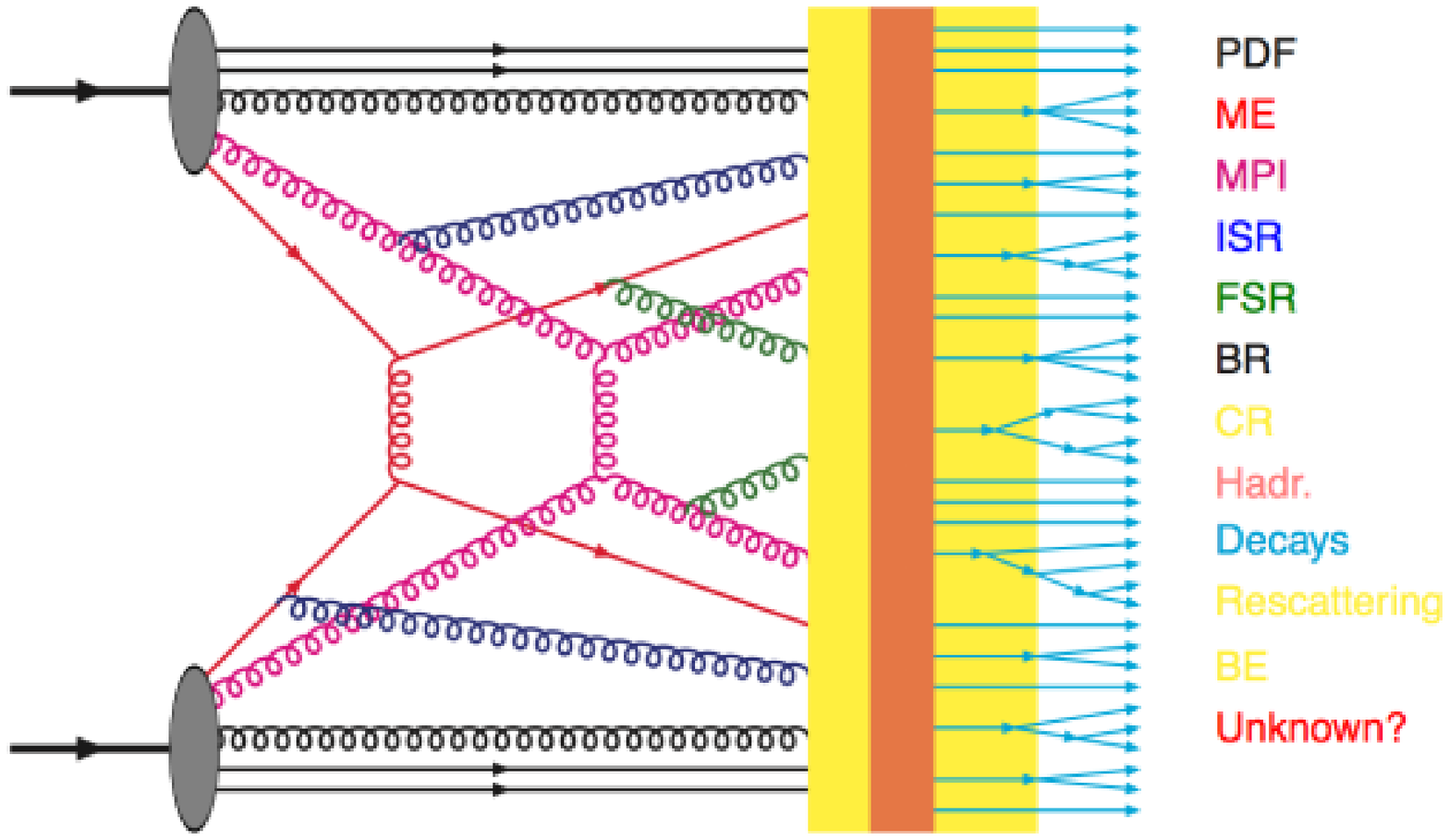}\\
\emph{\textbf{Fig.1}} {\emph{All parts of high-energy collisions.  }}
\end{center}

In Fig.2 is shown the structure of the putting together of presented processes for the concrete simulation in Angantyr model. 

\begin{center}
 \includegraphics[width=1\textwidth]{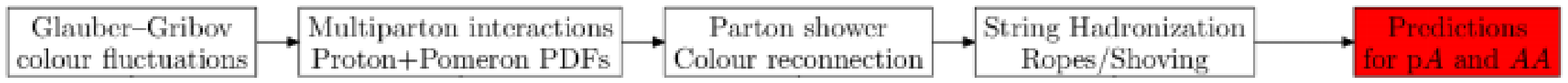}\\
 \vspace*{5mm}
\emph{\textbf{Fig.2}} {\emph{The illustration of each separate part of a normal PYTHIA8 simulation in order to make predictions for heavy ion collisions.  }}
\end{center}

\par
	We'll present each part separately in detail.\\
	\section{Models of nucleus-nucleus interactions}
\subsection{Glauber model of the nucleus-nucleus interactions with Gribov corrections}
One of the models describing the characteristics of the nucleus-nucleus interactions connected with the calculations of the number of interacting nucleons and binary NN collisions is the Glauber model \cite{1.}. This formalism is based on the eikonal approximation in impact parameter space, where the incident particle sequentially interacts with the nucleons of the target nucleus through multiple sub-collisions. The total hadronic cross section of two nuclei $A$ and $B$ is given by a $(2A +2B +1)$ - integral 

\begin{eqnarray*}
\sigma_{\text{AB}} = \int{d^2 b}\int{d^2 s_1^{\text{A}} \dots d^2 s_{\text{A}}^{\text{A}} d^2 s_1^{\text{B}} \dots d^2 s_{\text{B}}^{\text{B}} } \times \\
T_{\text{A}} s_1^{\text{A}} \dots T_{\text{A}} s_{\text{A}}^{\text{A}} T_{\text{B}} s_1^{\text{B}} \dots T_{\text{B}} s_{\text{B}}^{\text{B}} \times \\
 \left\{1 - \prod_{j=1}^{\text{B}}\prod_{i=1}^{\text{A}}{ \left[ 1 - \sigma(\text{b} -s_i^{\text{A}} + s_j^{\text{B}} ) \right] } \right\},
\end{eqnarray*}
where $b$ is impact parameter and $s$ denotes a position in the transverse plane. The interaction probability $\sigma(s)$ is normalized and gives the nucleon–nucleon inelastic cross section $\sigma_{\text{NN}} = \int{d^2s \sigma(s)}$. $T_{\text{A}}(b) = \int{d\zeta \rho_{\text{A}}(b, \zeta)}$ is nuclear thickness function, which describes the transverse nucleon density $\rho$ by integrating the nuclear density  along the longitudinal direction $(z)$.
	The importance of including diffractive excitation at very high energies was pointed out by Gribov \cite{2.} according to additional screening. Scatterings of diffractive nature for inelastic diffractive processes have been encoded into an event-by-event variation of $\sigma_{\text{NN}}$ given by the following probability distribution
\begin{equation*}
P_{\sigma}\left(\sigma_{\text{NN}}\right)=C\frac{\sigma_{\text{NN}}}{\sigma_{\text{NN}} + \sigma_0}\exp^{-\left( \frac{\sigma_{\text{NN}} - \sigma_0}{\sigma_0^{\Omega}} \right) },
\end{equation*}
where $\sigma_0$ denotes the mean $\sigma_{\text{NN}}$ value, and $\Omega$ is width. The normalization $C$ is computed from the provided input (mean $\sigma_{\text{NN}}$ and $\Omega$) requiring $\int{\sigma P d \sigma}/\int{P d \sigma} = \sigma_0$, with the dispersion given by the ratio of inelastic difraction over elastic cross section at $t = 0$ (zero momentum exchange).
There are three possibilities for the diffractive interaction:
\begin{enumerate}
\item single diffractive excitation of the target nucleon and process have different final states;
\item single diffractive excitation of the projectile and process have different final states;
\item double diffractive excitation and process have different final states;
\item central-diffractive events with the characteristic double-gap topology due to double-Pomeron exchange interpreted as a color neutral gluon system and process have different final states. The diffractive events mediated by a Pomeron exchange \cite{3.};
\item non-diffractive inelastic scattering with the exchange that carry color (not gluon pair as pomeron). The phase-space of the final state of non-diffractive interactions is typically filled by particles with suppressed rapidity gaps.
\end{enumerate}
All components of the total cross section including elastic and inelastic cross sections from Regge fits to data in hadronic pp collisions with one Pomeron and one Reggeon term are expressed by formula
\begin{equation*}
\sigma_{\text{TOT}}^{\text{pp}}(s) = \left( 21.70 s^{0.0808} + 56.08 s^{-0.4525} \right) \text{mb}
\end{equation*}
with s in units of $GeV^2$. The inelastic cross section is
\begin{equation*}
\sigma_{\text{INEL}}(s) = \sigma_{\text{TOT}}(s) - \sigma_{\text{EL}}(s)
\end{equation*}
with $\sigma_{\text{EL}}^{\text{pp}}(s) = \frac{\left( \sigma_{\text{TOT}}^{\text{pp}}(s) \right)^2 }{16\pi B_{\text{EL}}^{\text{pp}}(s)}$, $B_{\text{EL}}^{\text{pp}}(s) = 5 + 4s^{0.0808}$. It includes single-diffractive (SD), double-diffractive (DD), central-diffractive (CD), and non-diffractive (ND) components
\begin{equation*}
\frac{d\sigma_{\text{SD}}^{\text{pp}\rightarrow X_p}(s)}{dt dM_x^2} = \frac{g_{3\mathbb{P}}}{16\pi} \frac{\beta^3_{p^\mathbb{P}}}{M_X^2} F_{\text{SD}}\left( M_X  \right) \exp\left( B_{\text{SD}}^{X_p} t \right),
\end{equation*}
\begin{equation*}
\frac{d\sigma_{\text{DD}}^{\text{pp}}(s)}{dt dM_1^2 dM_2^2} = \frac{g_{3\mathbb{P}}}{16\pi} \frac{\beta^3_{p^\mathbb{P}}}{M_1^2 M_2^2} F_{\text{DD}}\left( M_1, M_2  \right) \exp\left( B_{\text{DD}} t \right),
\end{equation*}
\begin{equation*}
\sigma_{\text{CD}}(s) = \sigma_{\text{CD}}(s_{ref}) \left( \frac{\ln\left( 0.06 s/s_0 \right)}{\ln\left( 0.06 s_{ref}/s_0 \right)} \right)^{3/2},
\end{equation*}
\begin{equation*}
\sigma_{\text{ND}}^{\text{pp}}(s) = \sigma_{\text{INEL}}^{\text{pp}}(s) - \int{\left( d\sigma_{\text{SD}}^{\text{pp}\rightarrow X_p}(s) + d\sigma_{\text{SD}}^{\text{pp}\rightarrow p_X}(s) + d\sigma_{\text{DD}}^{\text{pp}}(s) + d\sigma_{\text{CD}}^{\text{pp}}(s) \right) } 
\end{equation*}
with difractive masses $(M_X, M_1, M_2)$, the Pomeron couplings $(g_{3\mathbb{P}}, \beta_{p^\mathbb{P}})$, the difractive slopes $(B_{\text{SD}}, B_{\text{DD}})$, the low-mass resonance-region enhancement and high-mass kinematical-limit suppression factors $(F_{\text{SD}}, F_{\text{DD}})$. The CD cross section at a fixed reference CM energy chosen to be $\sqrt{s_{ref}}=2\text{TeV}$ by default and $\sqrt{s_0} = 1\text{TeV}$.
The formation of inelastic intermediate states was taken later into account in \cite{4.} and presented by several simultaneous interactions in a sequence $p_{\bot1} \rightarrow p_{\bot2} \rightarrow p_{\bot3} \rightarrow p_{\bot4}$ $(p_{\bot} = \hat{t}\hat{u}/\hat{s})$, Fig.3
\begin{center}
 \includegraphics[width=0.4\textwidth]{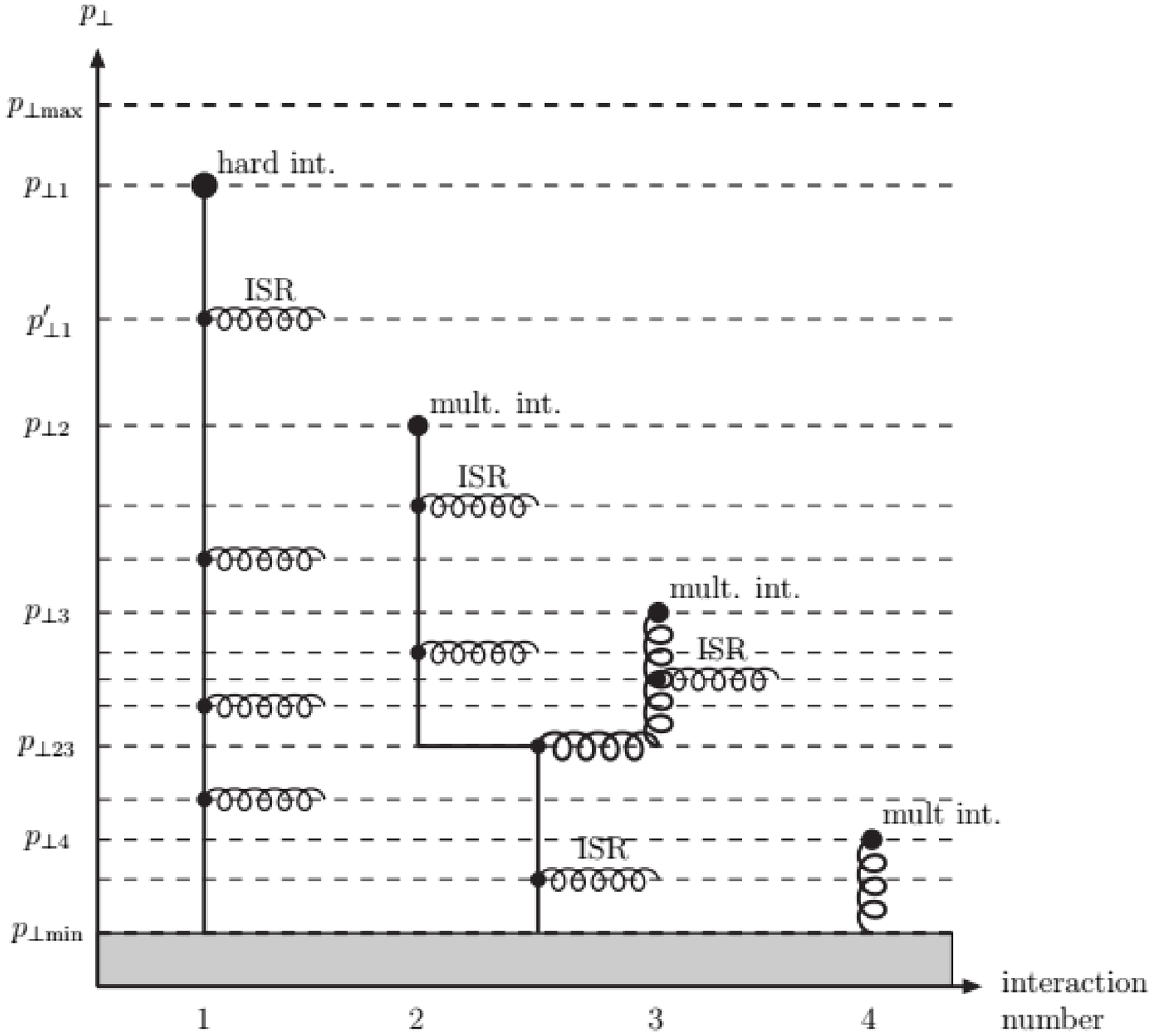}\\
\emph{\textbf{Fig.3}} {\emph{Schematic figure illustrating one incoming hadron in an event with a hard interaction occurring at p$\perp$1 and three further interactions at successively lower p$\perp$ scales, 
associated with initial-state radiation, and further with the possibility
of two interacting partons (2 and 3 here) having a common ancestor in the parton showers, from \cite{4.}.}}
\end{center}

Treatment of ISR and FSR is connected with replacement of each simple $2 \rightarrow 2$
interaction by more complicated $2 \rightarrow n$ process, $n \geq 2$, where additional partons are produced by ISR or FSR. The ISR and FSR algorithms are based on the dipole-style $p_{\bot}$-ordered evolution of hardness, since for gluon exchange processes 
$qq^{'} \rightarrow qq^{'}$, $qg \rightarrow qg$, $gg \rightarrow gg$ dominate the cross section. Integrated over the kinematically allowed range of $z$ and expressed as a differential branching probability per unit evolution time, the shower evolution with FSR and ISR is:

\begin{equation*}
\frac{d \cal{P}_{\text{FSR}}}{dp^2_{\bot}} = \frac{1}{p^2_{\bot}}\int{dz\frac{\alpha_s}{2\pi}P(z)},
\end{equation*}

\begin{equation*}
\frac{d \cal{P}_{\text{ISR}}}{dp^2_{\bot}} = \frac{1}{p^2_{\bot}}\int{dz\frac{\alpha_s}{2\pi}P(z)\frac{f'\left( x/z, p^2_{\bot} \right) }{zf(x, p^2_{\bot})}},
\end{equation*}
with $P(z)$: 

\begin{eqnarray*}
P_{q\rightarrow qg}(z) = C_F \frac{1+z^2}{1-z} ,\\
P_{g\rightarrow gg}(z) = C_A \frac{(1-z(1-z))^2 }{z(1-z)} , \\
P_{g\rightarrow q\bar{q}}(z) = T_R(z^2 + (1-z)^2) ,\\
\end{eqnarray*}
with $C_F = 4/3$, $C_A = N_c = 3$, and $T_R = 1/2$,  multipliet by $N_t$, if summing over all contributing quark flavors, for QCD with $z = x/x^{'}$ for ISR defined so $x<x^{'}$, and

\begin{equation*}
p_{\bot} = p_{\bot \text{evol}} = 
\begin{cases}
    (1-z)Q^2,& \text{ISR}\\
    z(1-z)Q^2,              & \text{FSR}
\end{cases}
\end{equation*}

with $Q_{\text{FSR}^2} = (p^2 - m_0^2)$ and $Q_{\text{ISR}^2} = (-p^2 + m_0^2)$ at $\alpha_s(M_Z)^{\text{pythia}} \sim 0.139 $.

\subsection{ Multiparton interactions}

All partonic sub-collisions are treated as separate QCD $2 \rightarrow 2$ scatterings. To eliminate divergences of the cross section at low $p_{\bot}$ was used a parameter $p_{\bot 0}$ which depends on the collision energy:

\begin{equation*}
\frac{d\sigma_{2 \rightarrow 2} }{dp_{\bot}^2} \propto \frac{\alpha_s^2(p_{\bot}^2)}{p_{\bot}^4} \rightarrow \frac{\alpha_s^2 \left( p_{\bot}^2 + p_{\bot 0}^2 \right)}{\left( p_{\bot}^2 + p_{\bot 0}^2 \right)^2}
\end{equation*}

Using assumption about the matter distribution in the colliding protons, the cross section gets a relative probability for each additional sub-scattering, presented in Fig.4.

\begin{center}
 \includegraphics[width=0.65\textwidth]{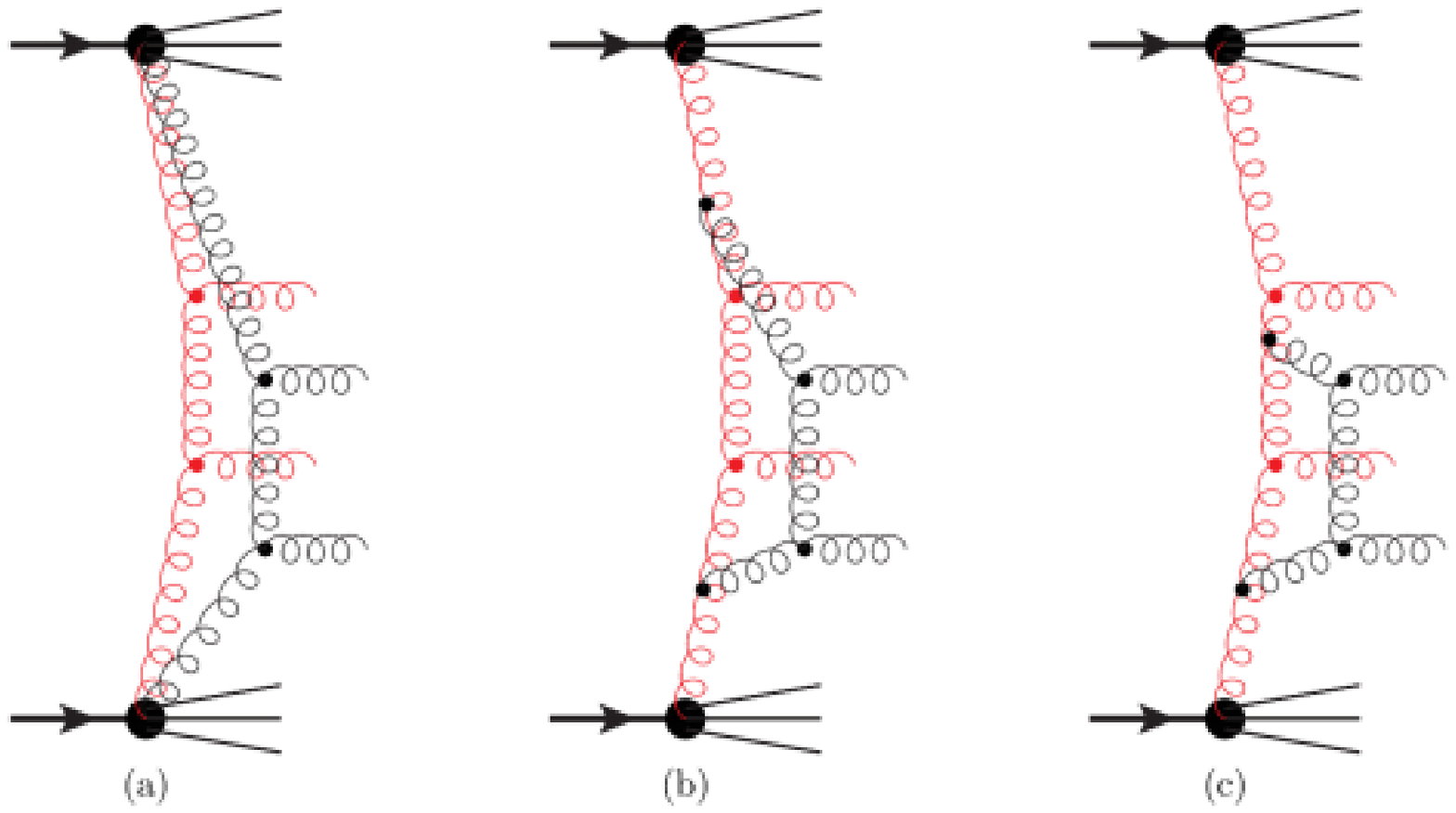}\\
\emph{\textbf{Fig.4}} {\emph{Schema of multi-parton interactions in a pp collision. Each gluon should be interpreted as having two colour lines, which after string hadronisation will contribute to the soft multiplicity, (a) the colour lines for both sub scatterings stretches out to the proton remnants, (b) and (c) the sub scattering is colour-connected inside the proton remnants, from \cite{5.}.}}
\end{center}

\subsection{Lund string model of hadronisation}
Recent precise measurements at the LHC show: 
\begin{itemize}
\item flow-like effects in pp and p-nucleus collisions, \cite{6.};
\item increasing strangeness and baryon rates in pp events with high multiplicity, \cite{7.};
\item enhanced transverse momenta and rising of angular correlations expansion for high mass particles in high energy collisions  allowing for rope formation, \cite{8.}.
\end{itemize}

These collective effects in nucleus collisions could possibly originate not from QGP formation but from non-thermal interactions between string-like colour fields with the dense configurations of confined QCD flux tubes with high string density. The Lund string model \cite{9.} described by a "massless relativistic string", presents flux tube with no transverse extension. Gluons are treated as point-like transverse excitations on the string \cite{10.}, stretched from a quark via the colour-ordered gluons to an antiquark. The connected in strings partons then evolve in a partonic cascade after freeze-out, and hadronise according to the Lund model and then the obtained hadrons form a secondary cascade as free particles. The probability for a final state is given by \cite{11.}.

\begin{equation*}
{d\cal{P}} \propto \exp(-bA) \times d\Omega
\end{equation*}
with $b$ - action for the relativistic string, $A$ – the space-time area (in units of the string tension $k$) covered by the string before its breakup into hadrons (Fig.5), the phase space $\Omega$ (in $1+1$ dimensions) is given by $d\Omega = \prod_{i=1}^n \left[ N d^2p_i\delta(p_i^2-m^2) \right]\delta^{(2)} \left( \sum{p_i - P_{\text{tot}}} \right) $.

\begin{center}
 \includegraphics[width=0.5\textwidth]{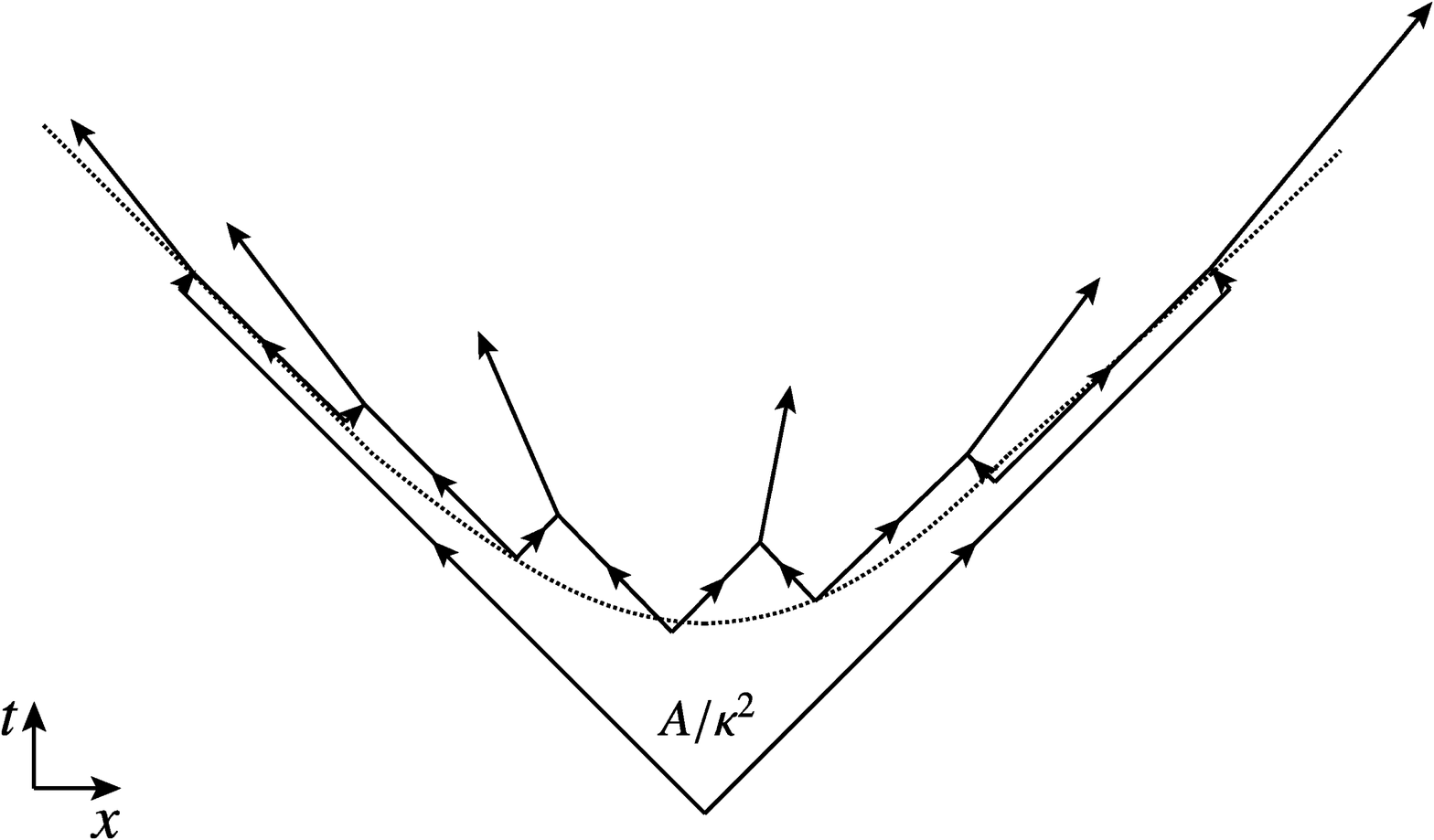}\\
\emph{\textbf{Fig.5}} {\emph{String between a quark and an antiquark in a $x - t$ diagram with $q\overline{q}$ pairs moving along light-like trajectories and produced around a hyperbola, to combine the hadrons, from \cite{12.}.}}
\end{center}

To each hadron is given a fraction $z$ of the light-cone momentum, determined by the distribution for single hadron with mass $m$, $f(z) = Nz^{\alpha} \exp{(-bm^2/z)}$ with three parameters $N$, $\alpha$, and $b$ determined by normalisation and by experiments.

\section{Results of calculations}
In the framework of PYTHIA8.3 program \cite{13.}, the inclusion of Angantyr model for heavy ions \cite{14.} gave us the opportunity to make calculations of p-Pb and p-Au heavy ion collisions at the energy of 5.02 TeV and 8.14 TeV. The advantage of this program is in the combining of several nucleon-nucleon collisions into one heavy ion collision, based on phenomenological treatment of a hadron as a vortex line in a colour superconducting medium, the consistent treatment of the central rapidity region with improvements of Glauber-like model where diffractive excitation processes are taken into account. 

Below, in Fig.6 the distributions of the number of events with respect to the transverse momentum $p_T$ and rapidity $Y$ for K-mesons on the left and $\Lambda$-baryon on the right at the energy of 5.02 TeV are presented.

\begin{center}
 \includegraphics[width=0.4\textwidth]{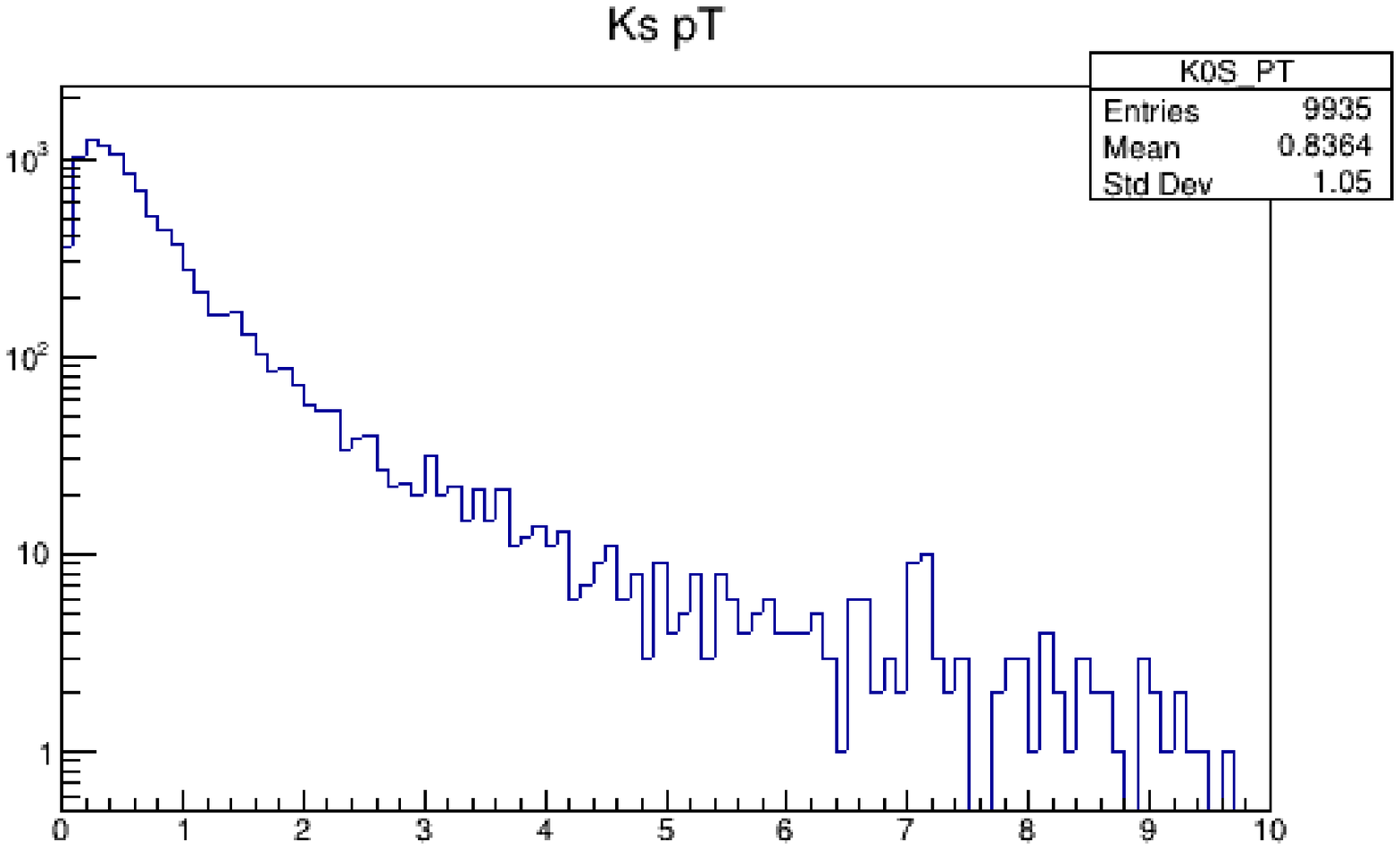}
  \includegraphics[width=0.4\textwidth]{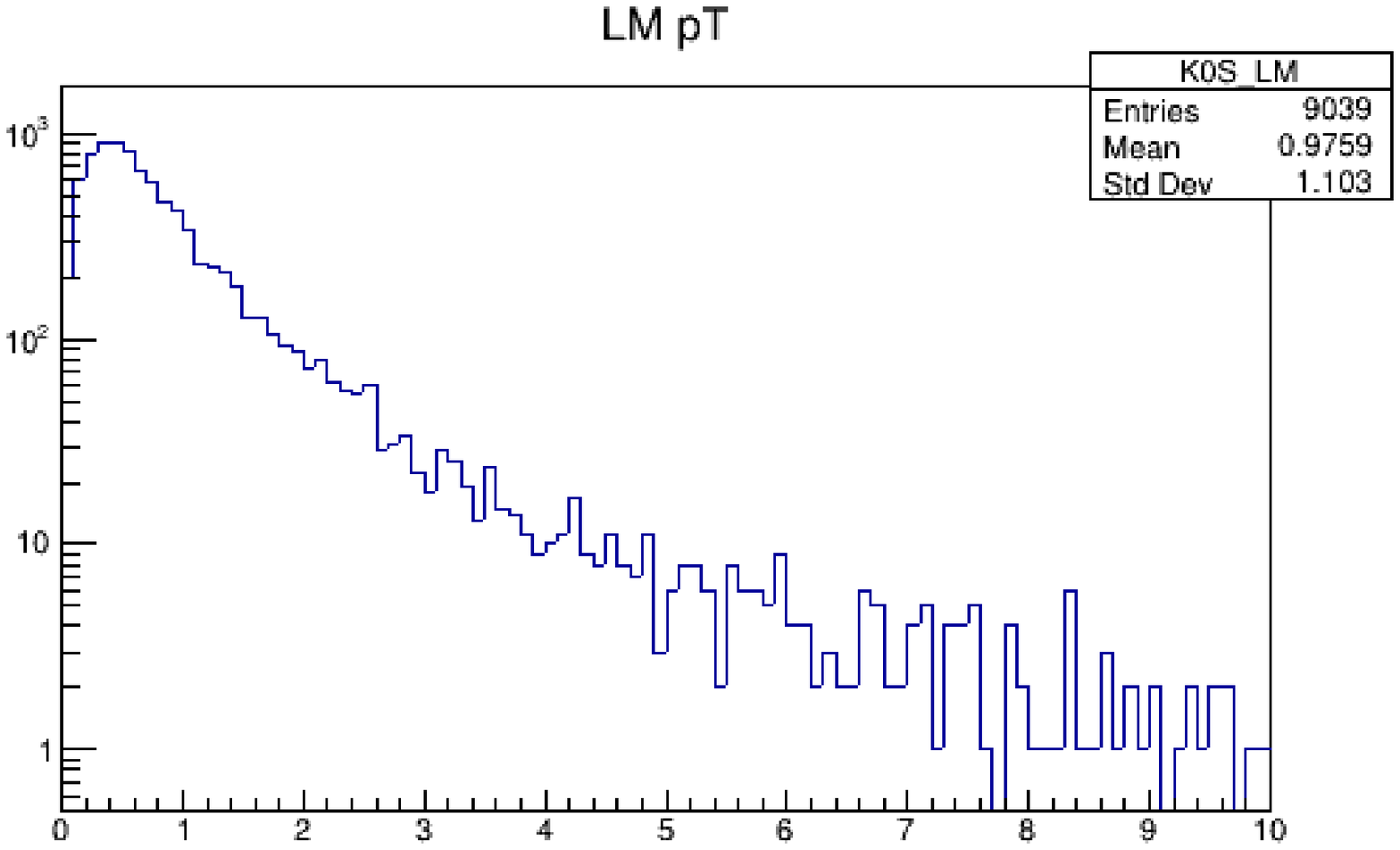}
   \includegraphics[width=0.4\textwidth]{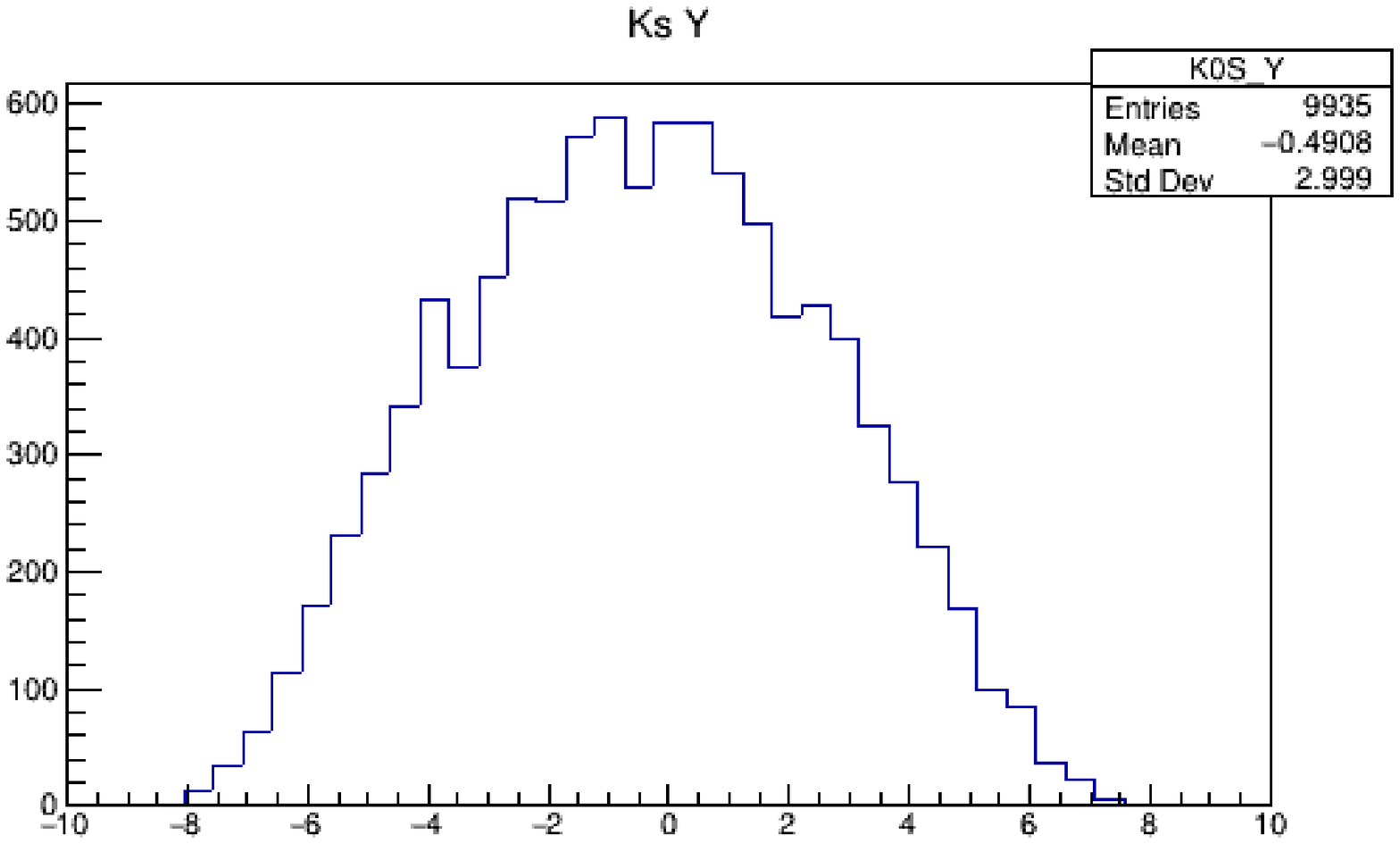}
    \includegraphics[width=0.4\textwidth]{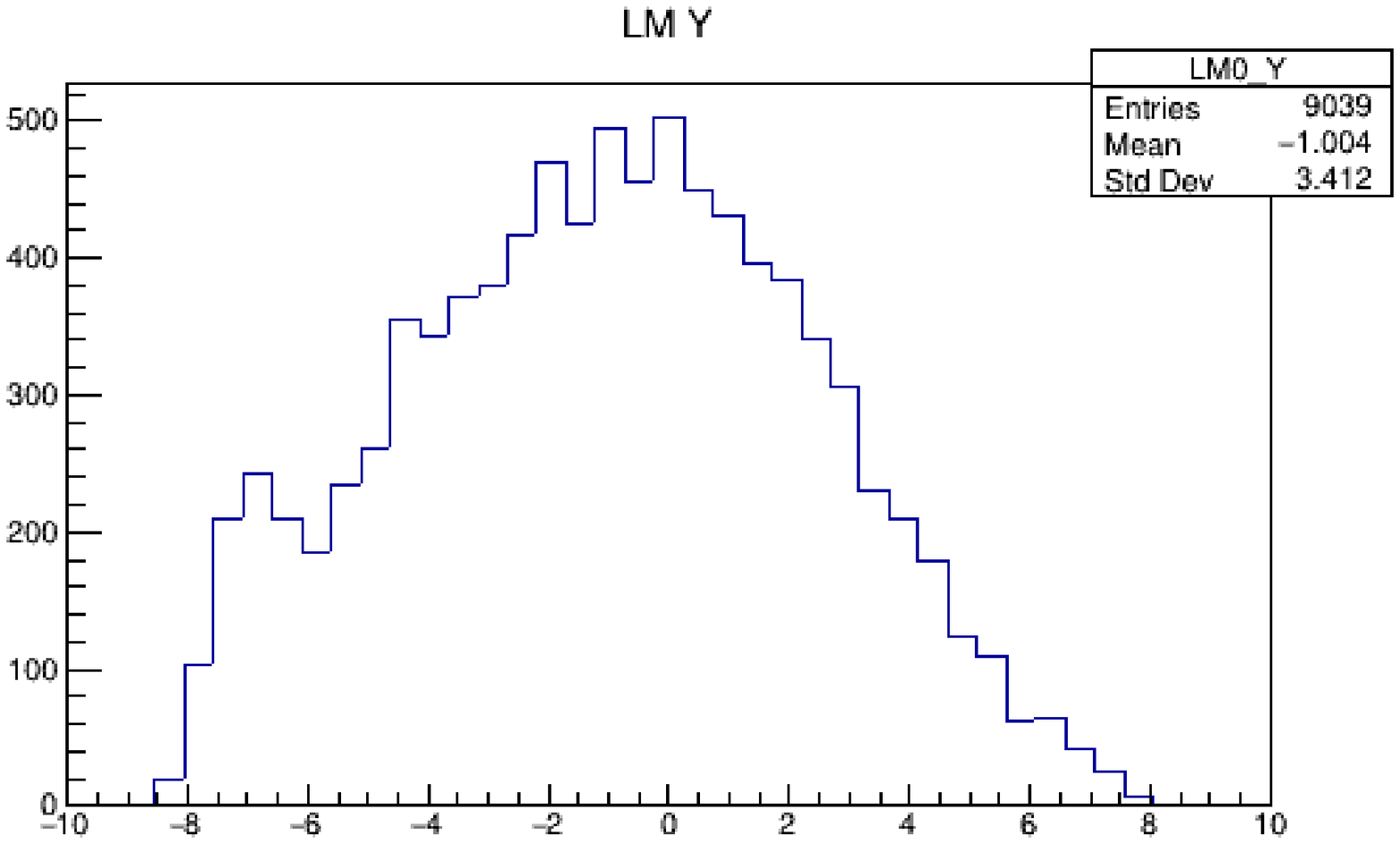}\\
\emph{\textbf{Fig.6}} {\emph{The entries with respect to the transverse momentum $p_T$ and rapidity $Y$ for K-mesons on the left side and for $\Lambda$-baryon on the right at the energy of 5.02 TeV.}}
\end{center}

We have considered the influence of impact parameter correlations on the production cross sections of p-Pb and p-Au processes to estimate the influence of hard and soft subprocesses on basic hadronic final-state properties in proton-ion collisions. Using these characteristics based on semi-hard multiparton interaction model we received the transverse momentum and rapidity distributions of K-meson and $\Lambda$-baryon at the energy of 8.14 TeV. The corresponding data for the transverse momentum and rapidity distributions of K-meson are presented in Fig.7 and Fig.8 respectively.

\begin{center}
 \includegraphics[width=0.4\textwidth]{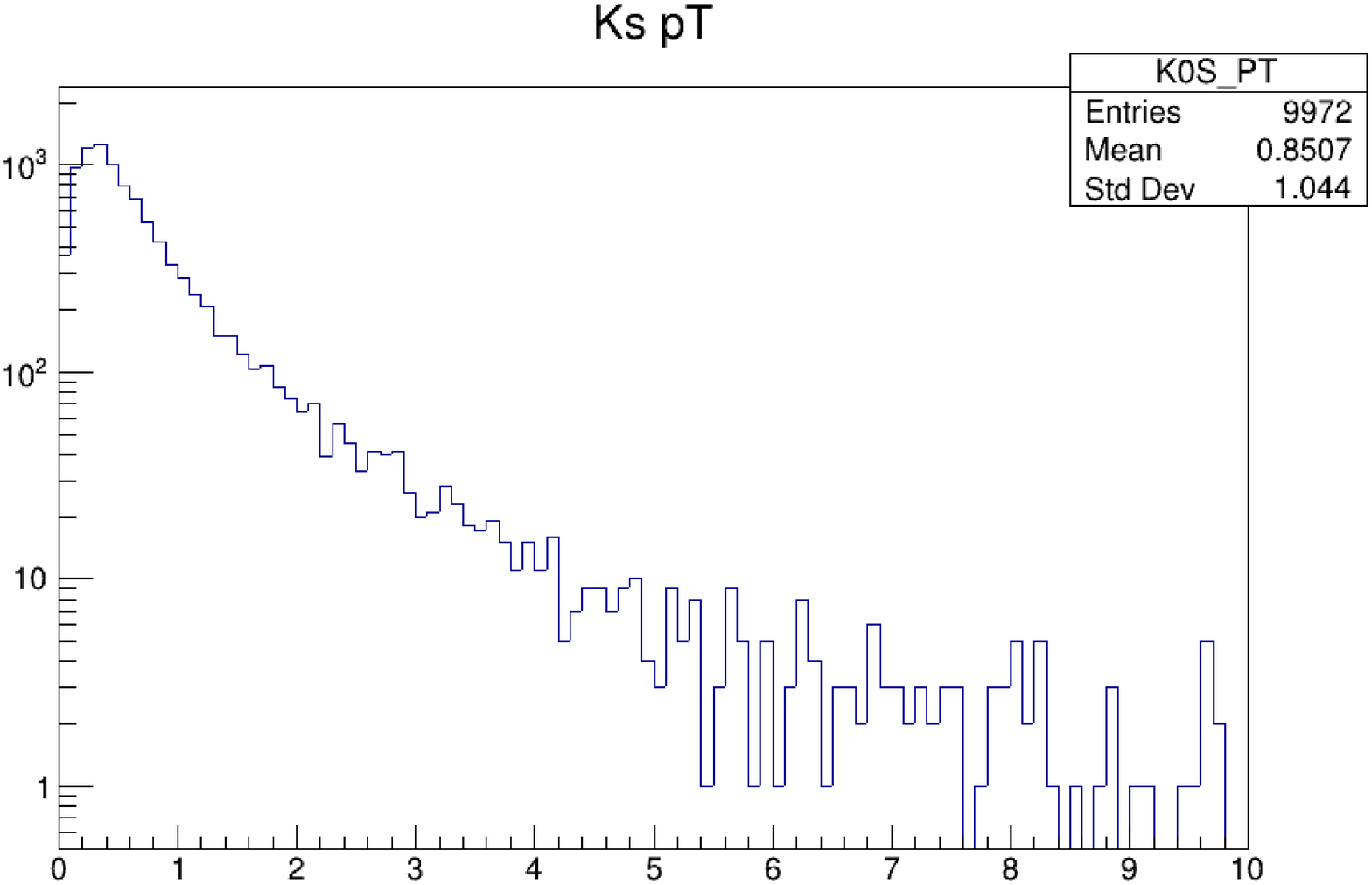}
 \includegraphics[width=0.4\textwidth]{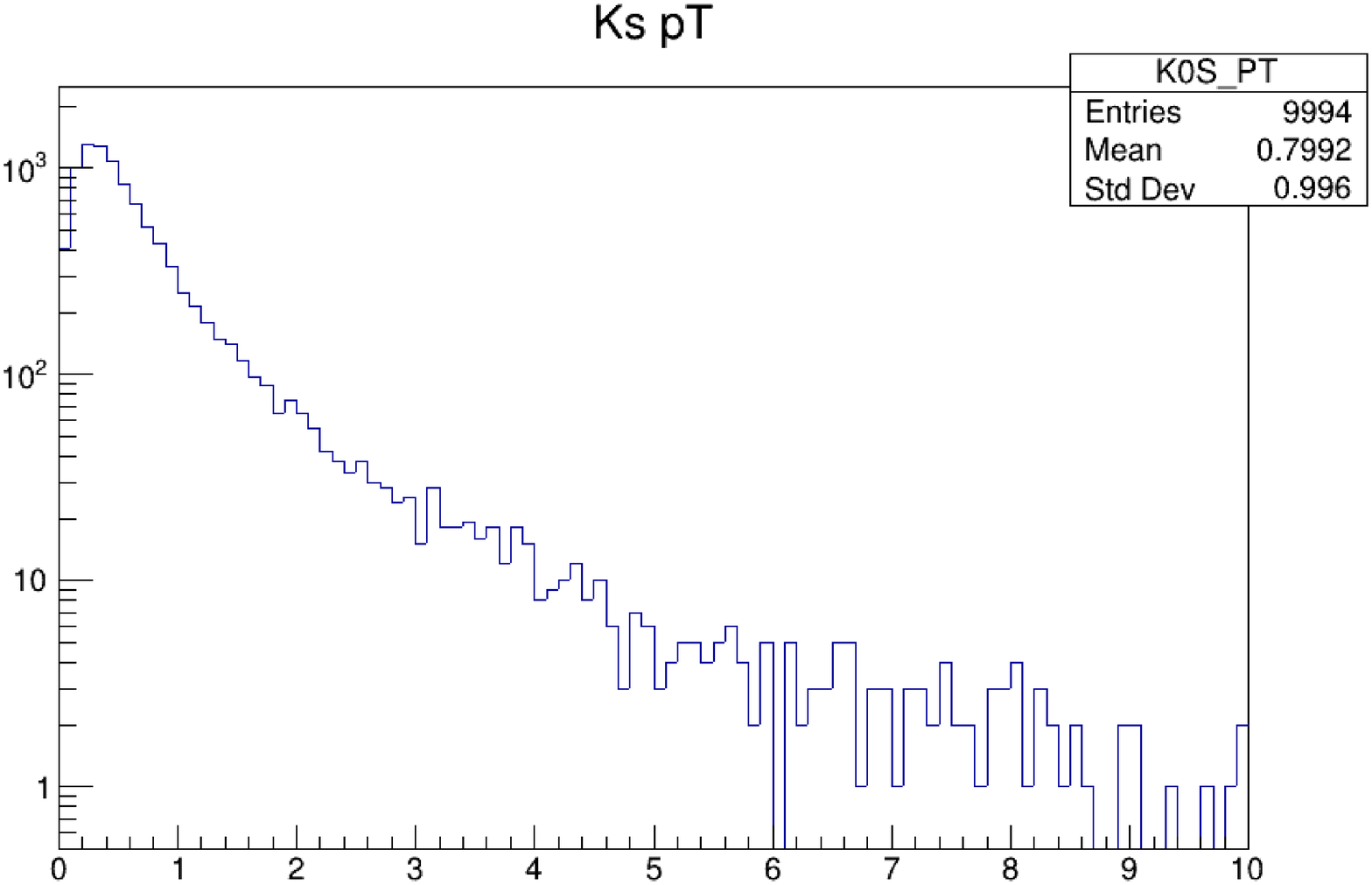}
 \includegraphics[width=0.4\textwidth]{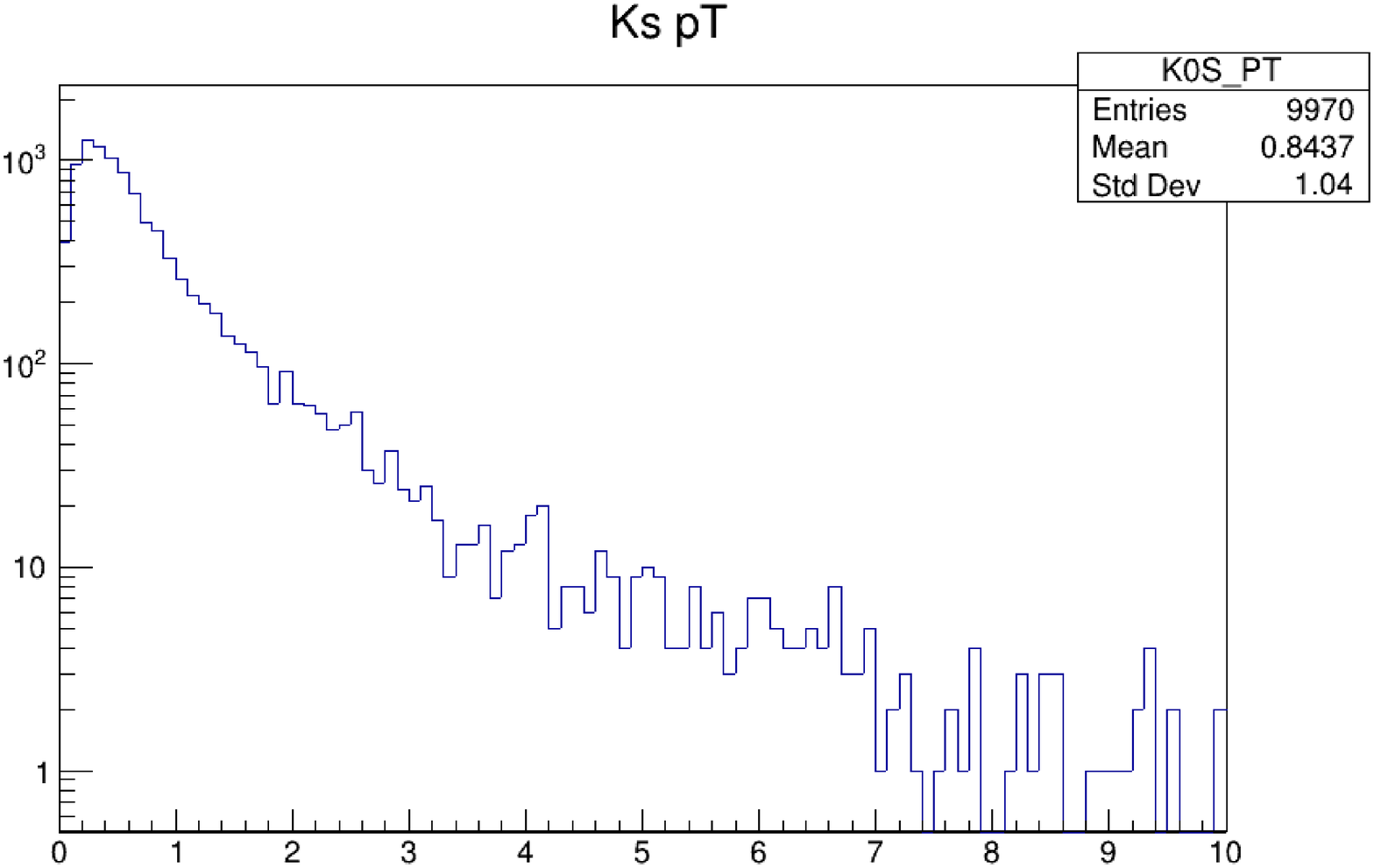}
 \includegraphics[width=0.4\textwidth]{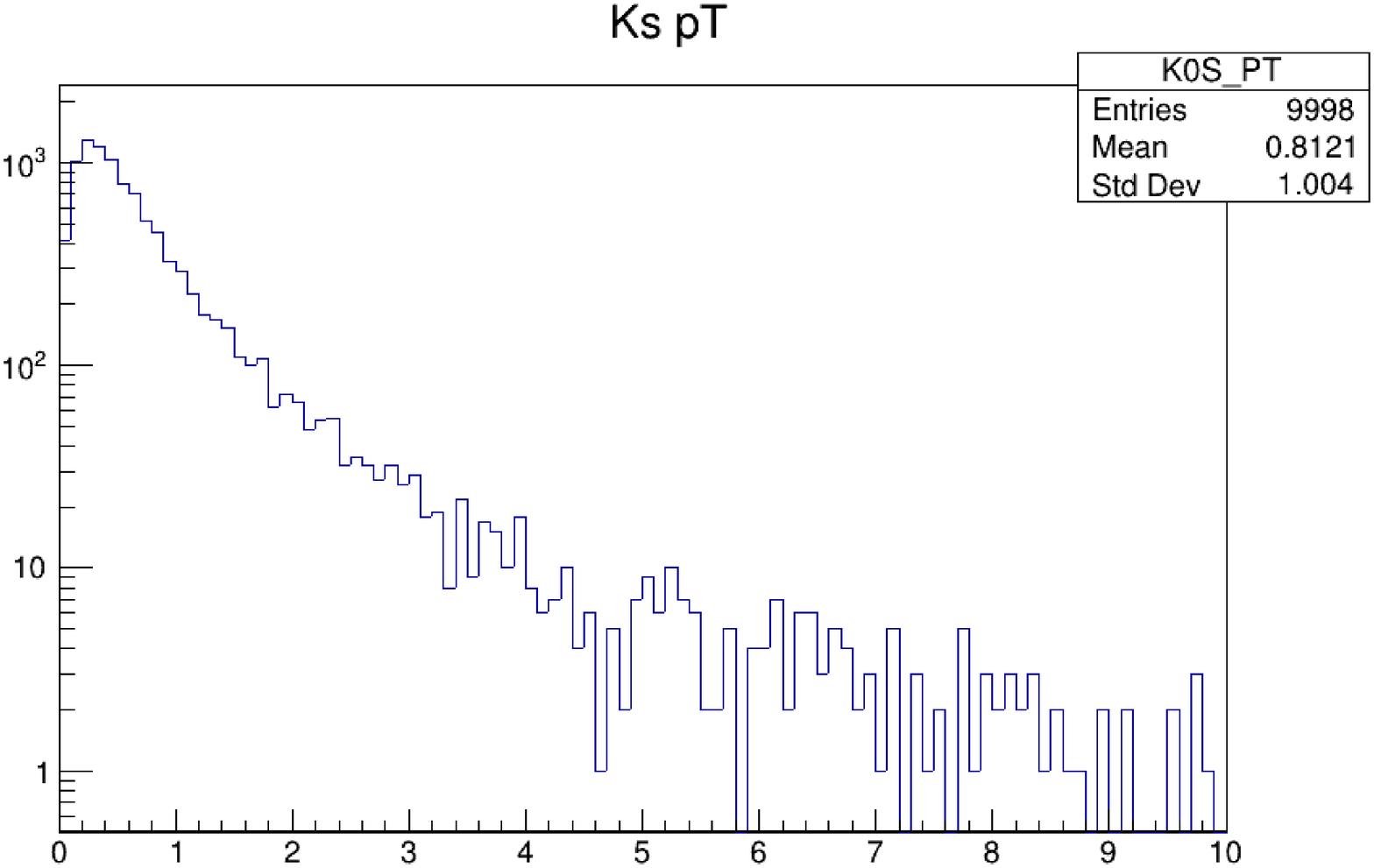}\\
\emph{\textbf{Fig.7}} {\emph{Left: Number of events for p-Au collision as a function of p$_T$ for Ks meson at b=0 up and b=0.5 down; Right: Number of events for p-Pb collision as a function of p$_T$ for Ks meson at $b=0$ up and $b=0.5$ down.}}
\end{center}
 
 \begin{center}
 \includegraphics[width=0.4\textwidth]{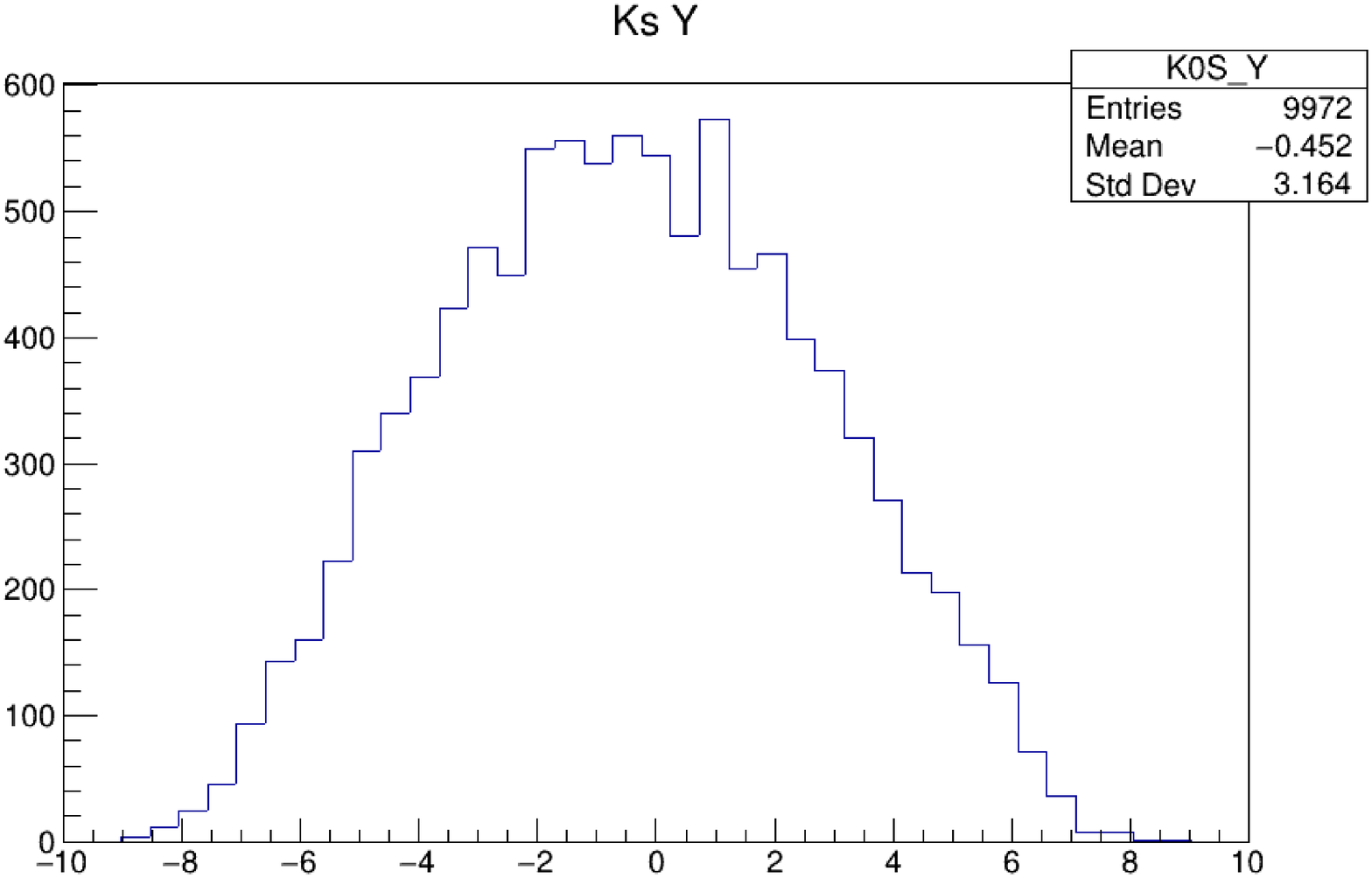}
 \includegraphics[width=0.4\textwidth]{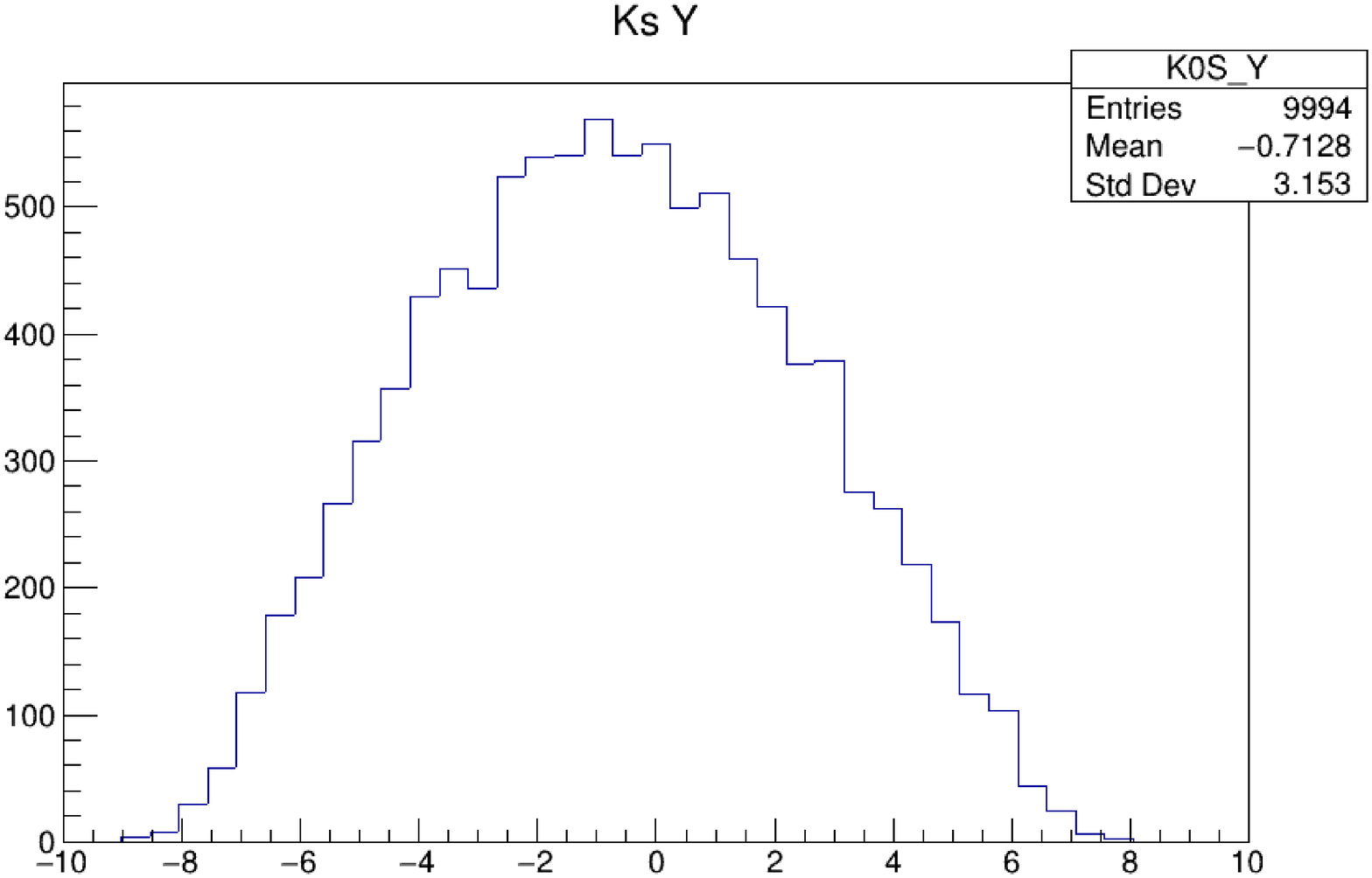}
 \includegraphics[width=0.4\textwidth]{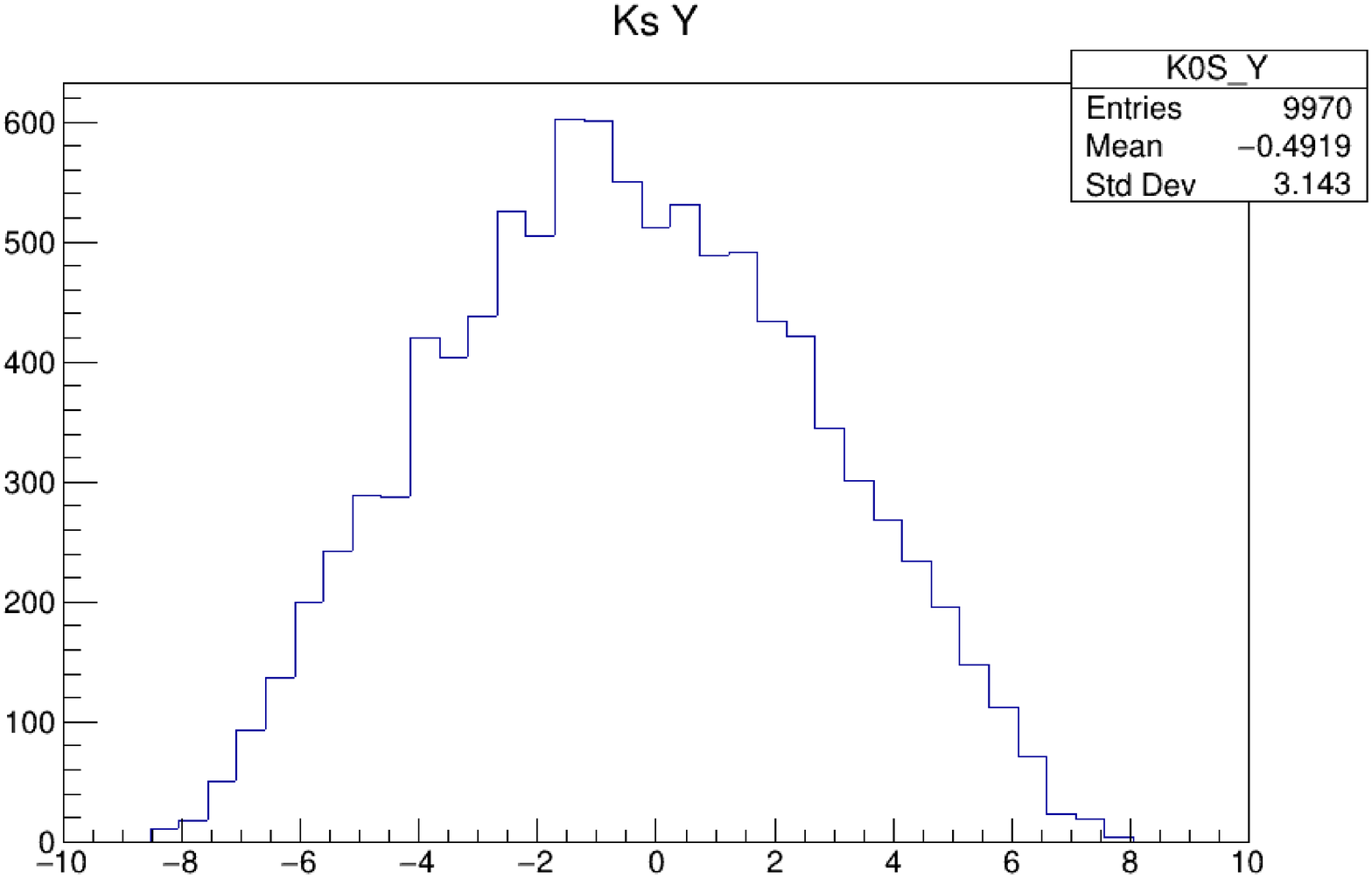}
 \includegraphics[width=0.4\textwidth]{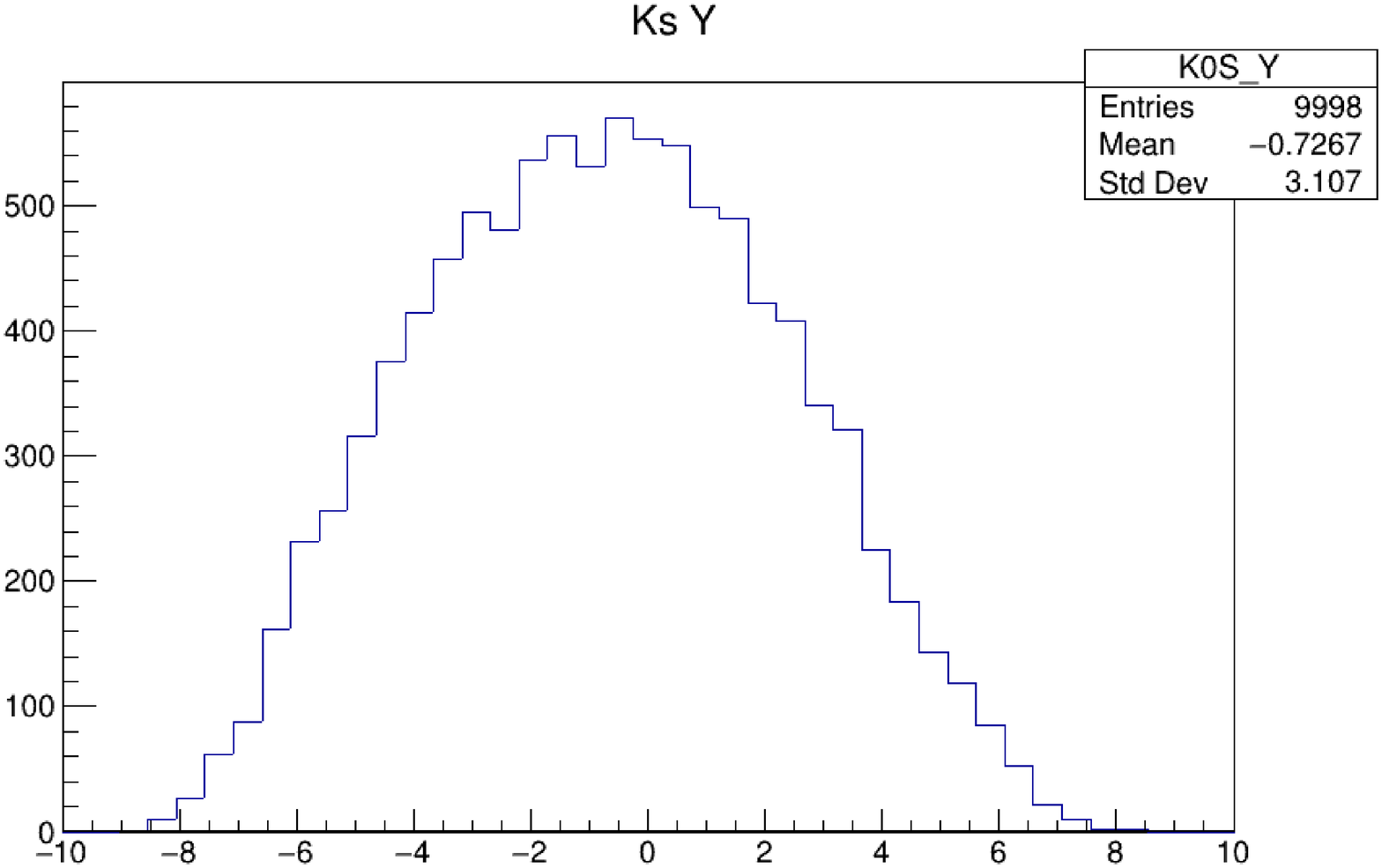}\\
\emph{\textbf{Fig.8}} {\emph{Left: Number of events for p-Au collision as a function of Y for Ks meson at b=0 up and b=0.5 down; Right: Number of events for p-Pb collision as a function of Y for Ks meson at $b=0$ up and $b=0.5$ down.}}
\end{center} 
 
The corresponding data for the transverse momentum and rapidity distributions of Lambda baryon are presented in Fig.9 and Fig.10 respectively.

\begin{center}
 \includegraphics[width=0.4\textwidth]{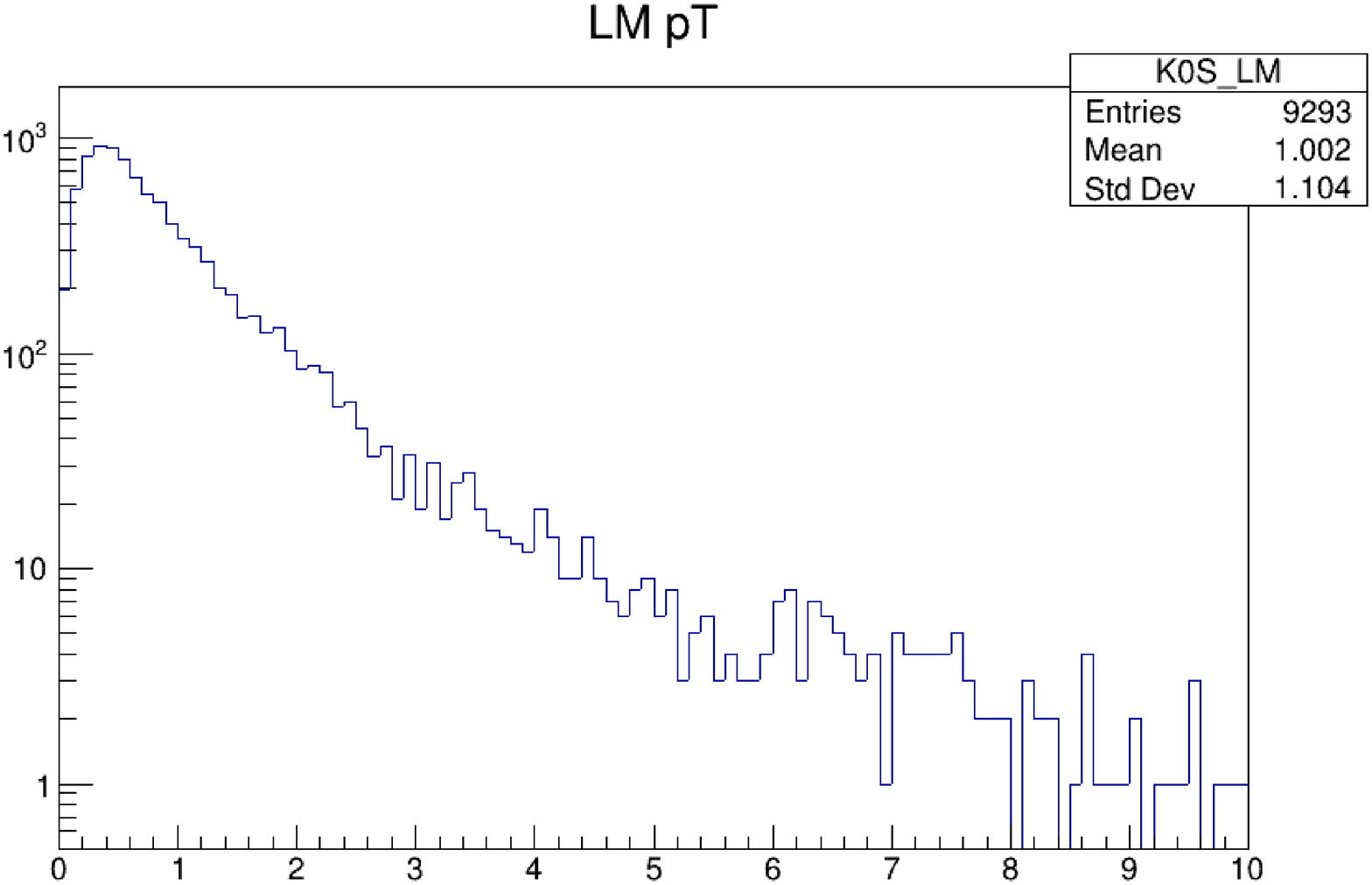}
 \includegraphics[width=0.4\textwidth]{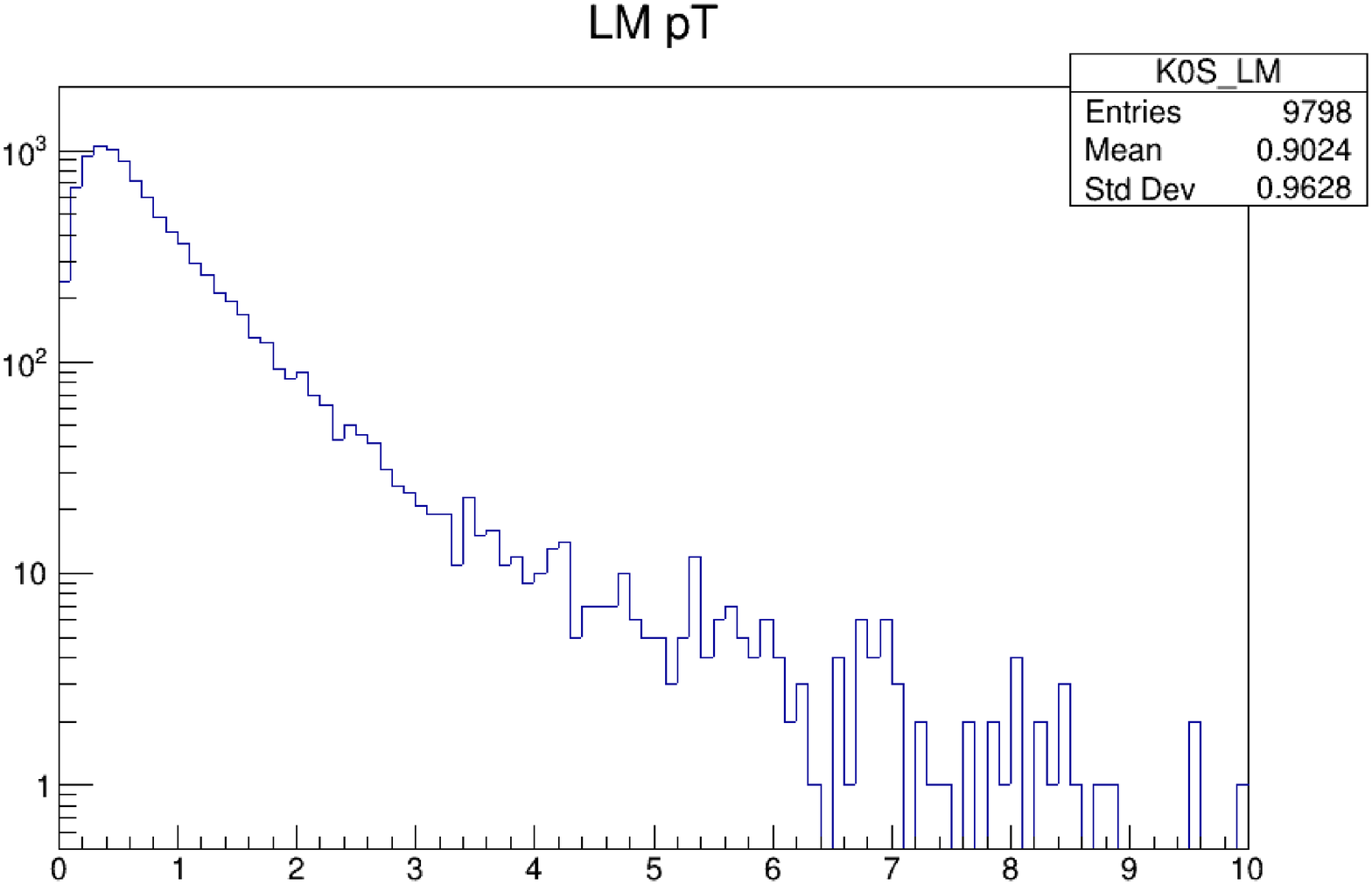}
 \includegraphics[width=0.4\textwidth]{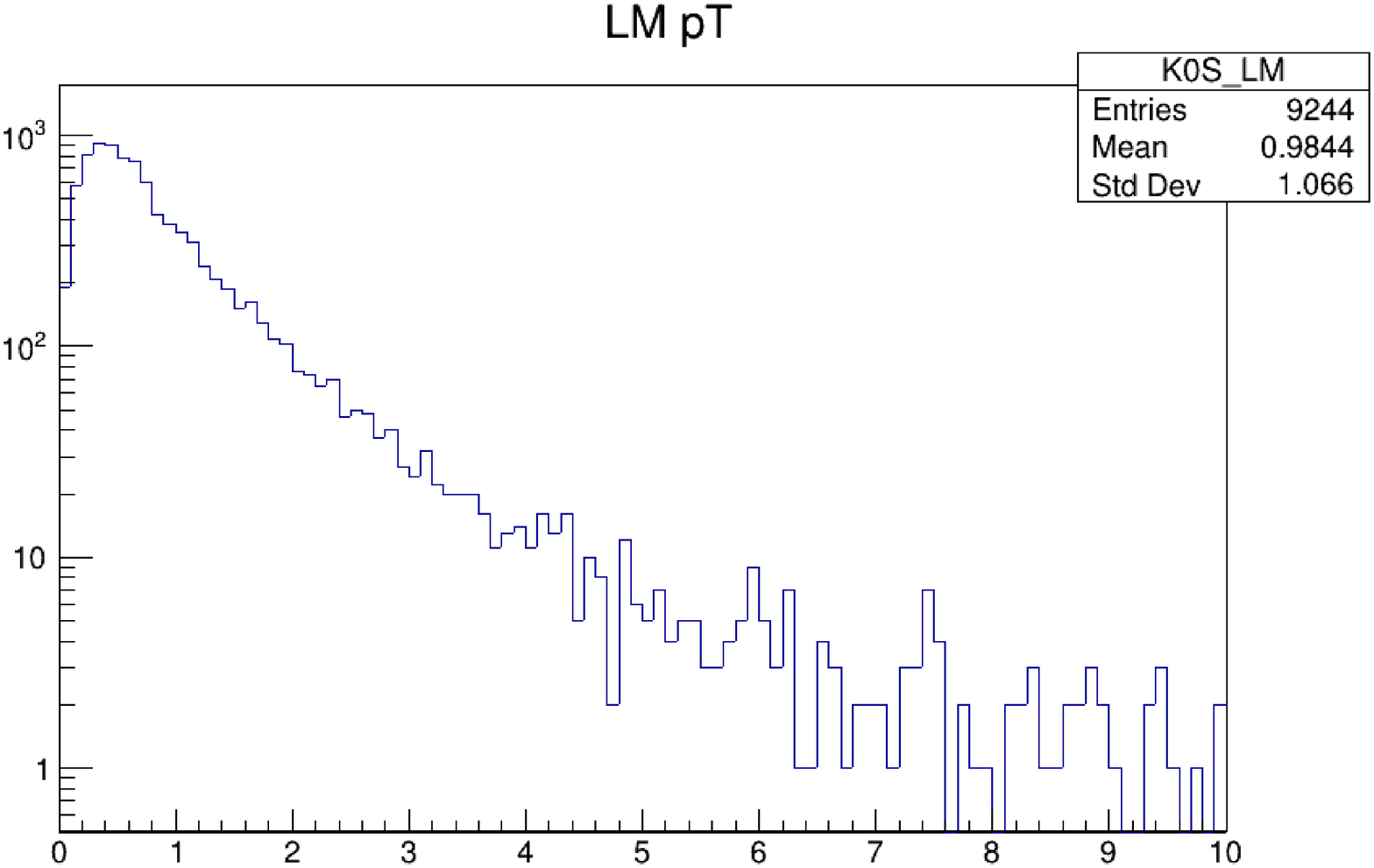}
 \includegraphics[width=0.4\textwidth]{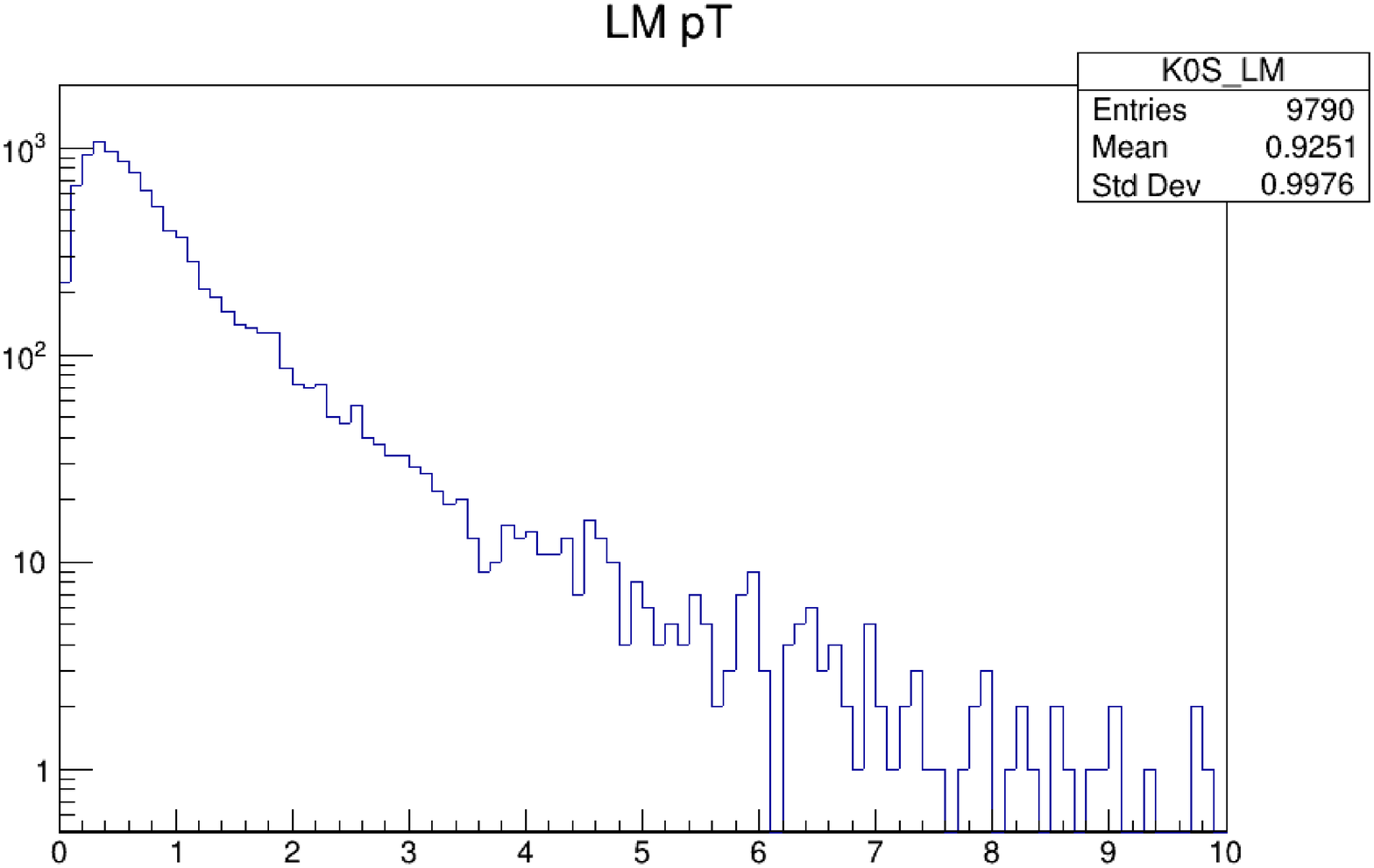}\\
 \emph{\textbf{Fig.9}} {\emph{Left: Number of events for p-Au collision as a function of $p_T$ for $\Lambda$-baryon at b=0 up and b=0.5 down; Right: Number of events for p-Pb collision as a function of $p_T$ for $\Lambda$-baryon at $b=0$ up and $b=0.5$ down.}}
 \end{center} 

\begin{center}
 \includegraphics[width=0.34\textwidth]{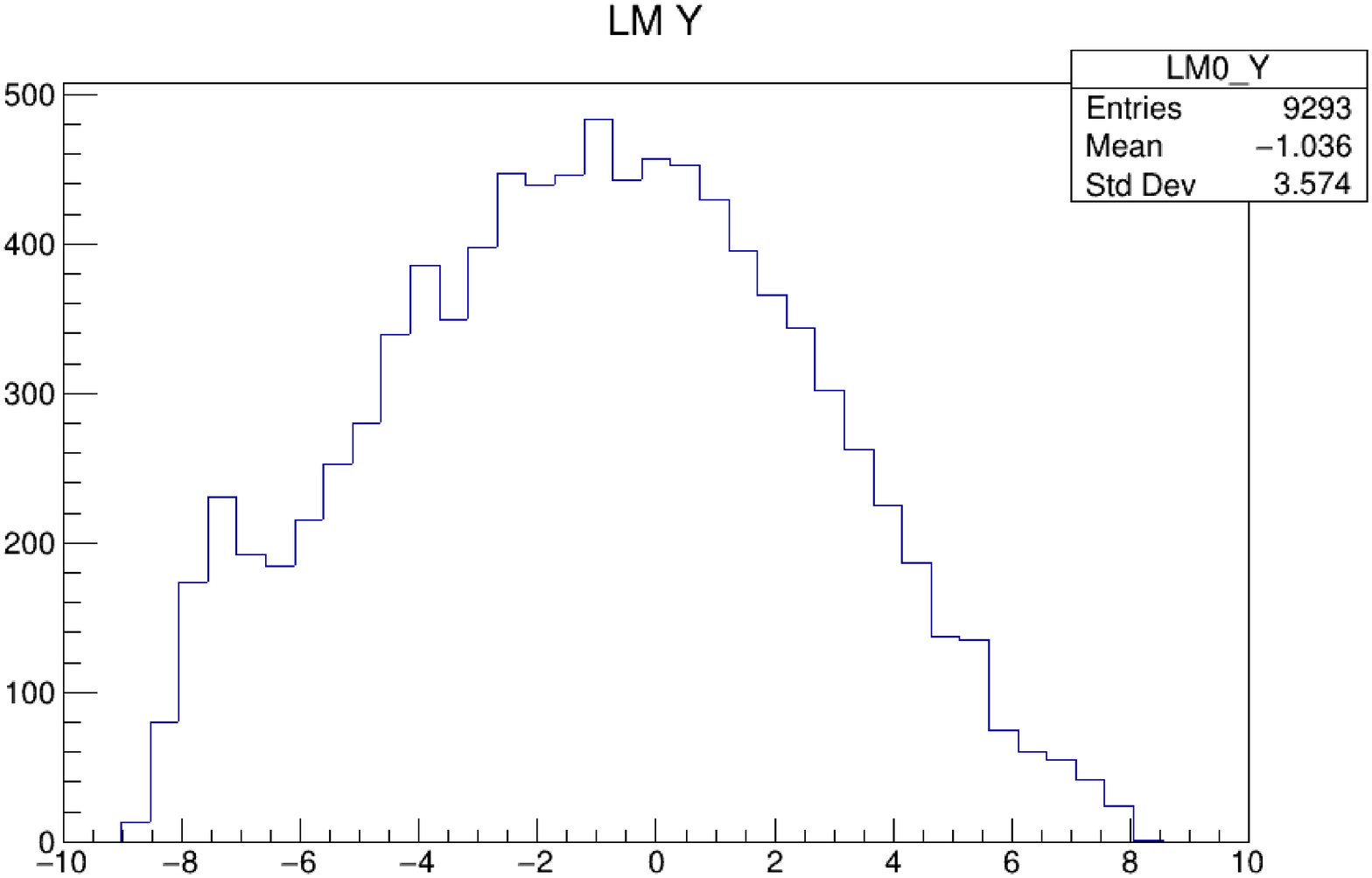}
 \includegraphics[width=0.34\textwidth]{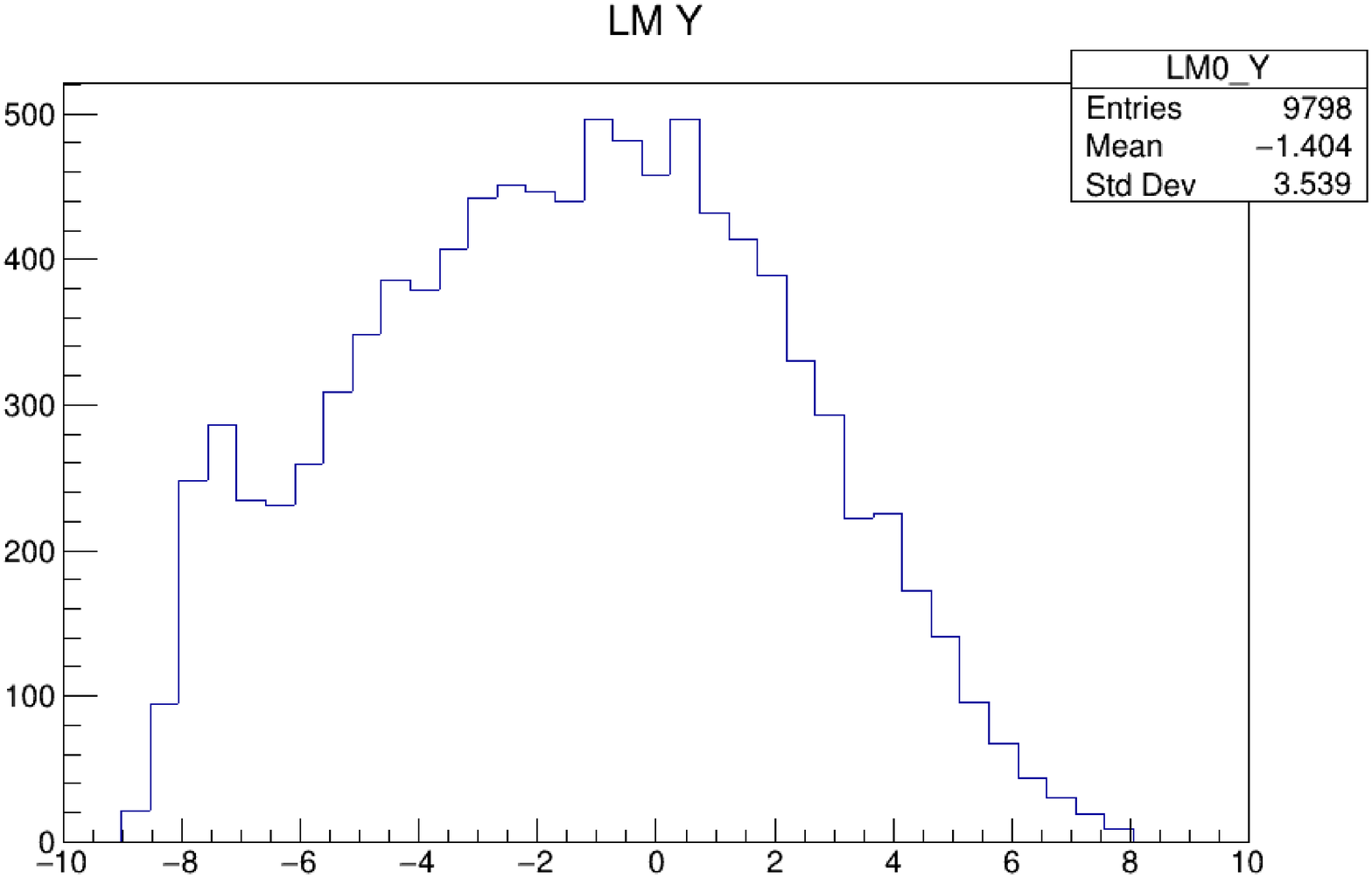}
 \includegraphics[width=0.34\textwidth]{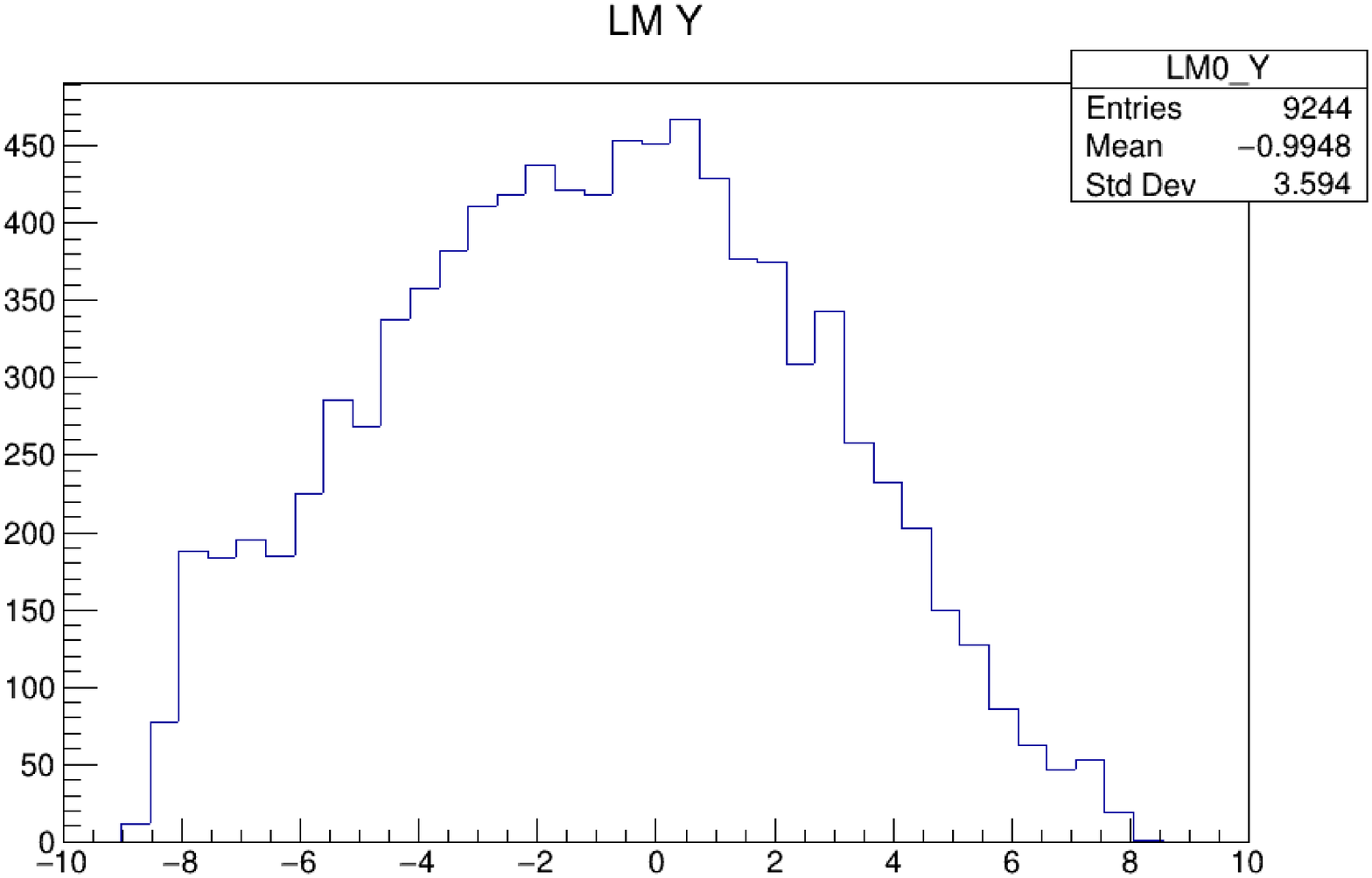}
 \includegraphics[width=0.34\textwidth]{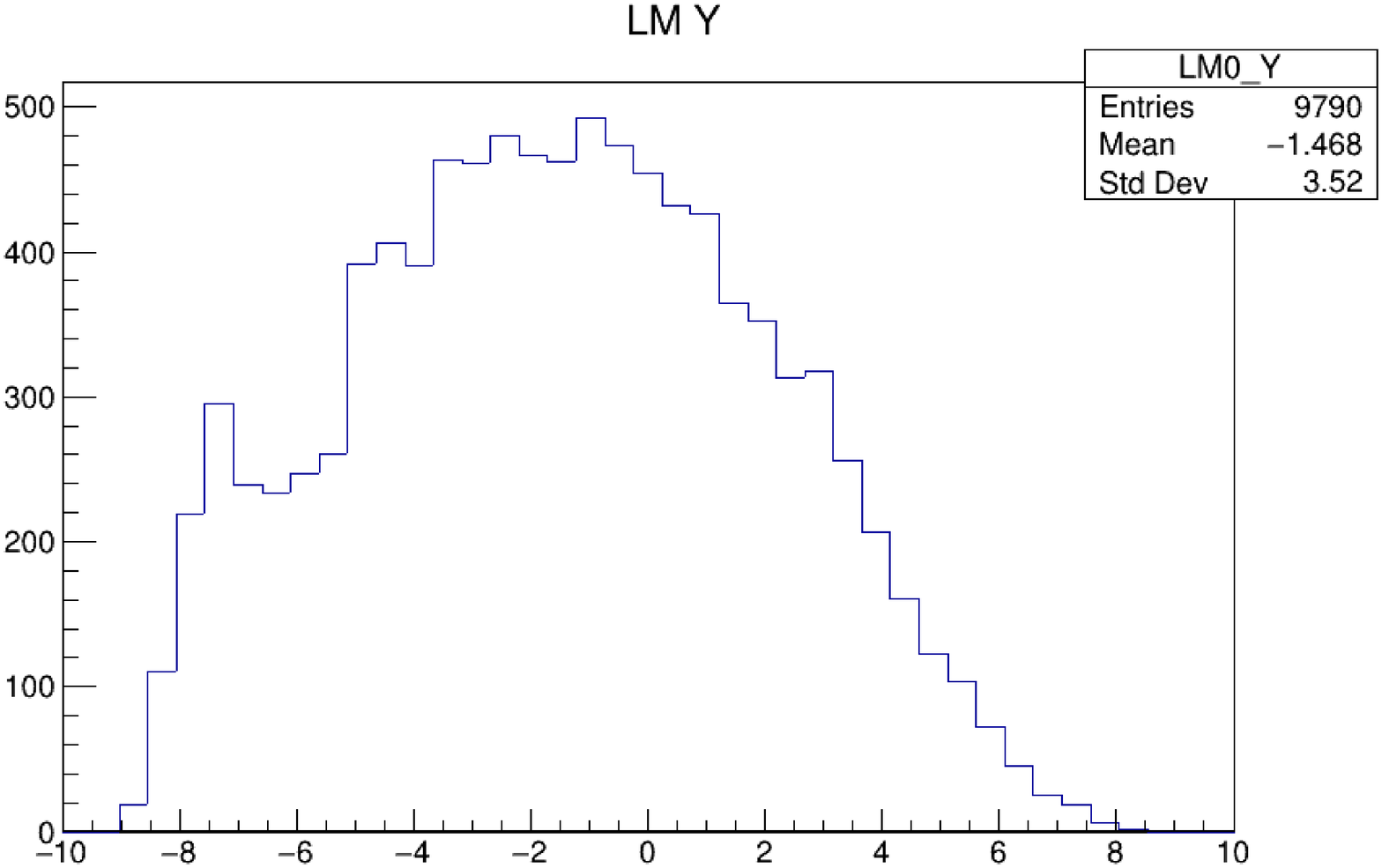}\\
 \emph{\textbf{Fig.10}} {\emph{Left: Number of events for p-Au collision as a function of $Y$ for $\Lambda$-baryon at $b=0$ up and $b=0.5$ down; Right: Number of events for p-Pb collision as a function of $Y$ for $\Lambda$-baryon at $b=0$ up and $b=0.5$ down.}}
\end{center} 

From the comparison of Fig.6 at 5.02 TeV and Fig.7-10 at 8.14 TeV we can see that the behavior of the functions is the same regardless of the collision energy with the exception of the number of events, which is natural.  In addition, the number of events for the K meson is always greater than for the $\Lambda$-baryon, which is also natural, since the baryon is heavier than the meson.

We also calculated multiplicity distributions for different impact parameter at the energy of 8 TeV. The results are in Fig.11.

\begin{center}
 \includegraphics[width=0.64\textwidth]{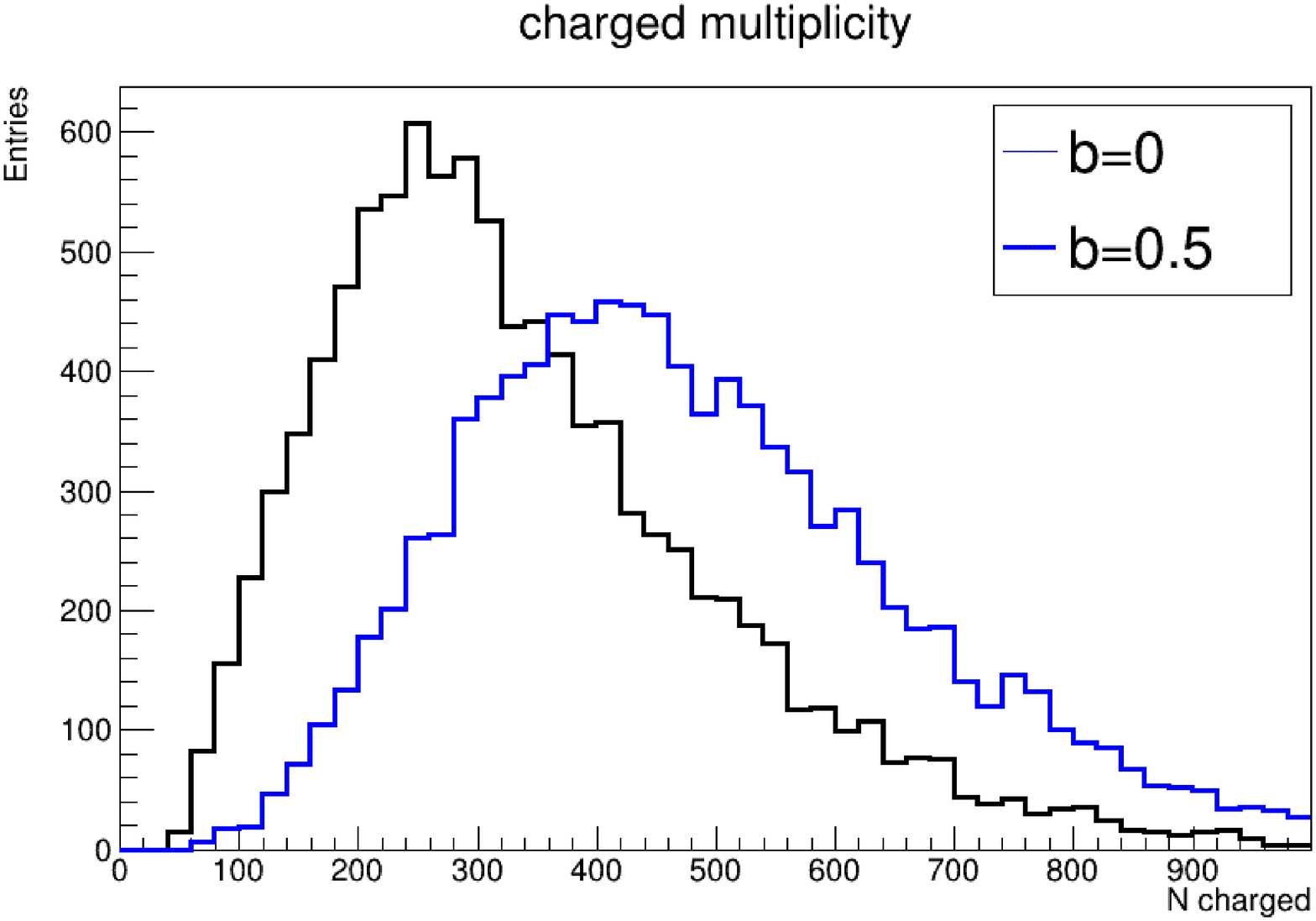} \\
  \emph{\textbf{Fig.11}} {\emph{Charged multiplicity distributions at 8 TeV for p-Au collisions for different impact parameters.}}
\end{center} 
From the comparison of the number of events for the charge multiplicity distribution, we see that the distribution peak decreases and shifts with an increase in the impact parameter from $0$ to $0.5$.

We investigated the influence of ISR and FSR on the production cross sections for p-Pb collisions at 8 TeV and presented the corresponding data in Table 1 and Table 2. 

\begin{center}
{\it\normalsize Table 1. The production cross sections for p-Pb collisions at 8 TeV \\ for inclusion of  ISR and FSR}\\
\vspace*{3mm}
\begin{tabular}{|c|c|c|}
\hline 
  Subprocess                           &      sigma $\pm$ delta    
                                           & (estimated) (mb)   \\
\hline
  g g $\rightarrow$  g g                           &   4.181e+01 & 2.275e-02 \\
  g g $\rightarrow$   $q\overline{q}$ (uds)                  &   8.301e-01 & 1.682e-01 \\
  $q g \rightarrow  q g$                           &   2.627e+01 & 2.855e-02 \\
  $q\overline{q}^{'}  \rightarrow q\overline{q}^{'}$      &   2.619e+00 & 8.534e-02 \\
  $q\overline{q} \rightarrow$  g g                        &   4.893e-02 & 1.117e+00 \\
   $q\overline{q} \rightarrow  {q}^{'} \overline{q}^{'}$ (uds)  &   1.034e-02 & 1.866e+00 \\
  g g $\rightarrow$  $c\overline{c}$                        &   2.810e-01 & 2.094e-01 
  \\
  g g $\rightarrow$  $b\overline{b}$                        &   1.987e-01 & 2.475e-01 \\
\hline
  sum                                  &   7.206e+01 & 1.724e-02 \\
\hline 
\end{tabular} 
\end{center}
\begin{center}
{\it\normalsize Table 2. The production cross sections for p-Pb collisions at 8 TeV \\ in case of disregard of ISR and FSR}\\
\vspace*{3mm}
\begin{tabular}{|c|c|c|}
\hline 
   Subprocess                           &      sigma $\pm$ delta  
                                           & (estimated) (mb)   \\
\hline 
  g g $\rightarrow$ g g                           &   4.250e+01 & 2.207e-02 \\
  g g $\rightarrow$   $q\overline{q}$ (uds)                 &   8.306e-01 & 1.714e-01 \\
 $ q g \rightarrow q g$                           &   2.452e+01 & 2.970e-02 \\
  $q\overline{q}^{'}  \rightarrow q\overline{q}^{'}$       &   3.001e+00 & 9.029e-02 \\
  $q\overline{q} \rightarrow$ g g                        &   4.946e-02 & 1.038e+00 \\
  g g $\rightarrow$  $c\overline{c}$                         &   3.305e-01 & 3.501e-01 \\
 $ q \overline{q} \rightarrow  c\overline{c}$               &   7.754e-03 & 1.403e+00 \\
  g g $\rightarrow$  $b\overline{b}$                        &   2.271e-01 & 2.655e-01 \\
 \hline 
  sum                                  &   7.147e+01 & 1.725e-02 \\
\hline 
\end{tabular} 
\end{center}

Comparison of the data in Table 1 and Table 2 leads us to the conclusion about an increase in the production cross section of p-Pb colllisions when taking into account ISR and FSR processes.

To study the effect of subprocesses during proton-proton interactions on the total production cross section, we performed calculations at 8 and 13 TeV, presented in Table 3 and Table 4. 

\begin{center}
{\it\normalsize Table 3. Subprocesses and corresponding value of production cross section\\ in pp collisions at 8 TeV.}\\
\vspace*{3mm}
\begin{tabular}{|c|c|c|}
\hline 
 Subprocess                           &      sigma $\pm$ delta &  (estimated) (mb)   \\
 \hline
 non-diffractive                      &   5.194e+01 & 0.000e+00 \\
 A B $\rightarrow \text{ A B elastic           }$        &   2.030e+01 & 0.000e+00 \\
 A B $\rightarrow \text{ X B single diffractive}  $      &   6.246e+00 & 0.000e+00 \\
 A B $\rightarrow \text{ A X single diffractive}   $     &   6.246e+00 & 0.000e+00 \\
 A B $\rightarrow \text{ X X double diffractive}    $    &   8.274e+00 & 0.000e+00 \\
 g g $\rightarrow \text{ g g                   }  $      &   1.999e-01 & 2.155e-02 \\
 g g $\rightarrow  q \overline{q}$ (uds)          &   0.000e+00 & 0.000e+00 \\
 $q g \rightarrow  q g $        &   1.594e-01 & 2.209e-02 \\
 $q \overline{q}^{'} \rightarrow  q \overline{q}^{'}  $        &   0.000e+00 & 0.000e+00 \\
 $q \overline{q} \rightarrow \text{ g g          }$        &   0.000e+00 & 0.000e+00 \\
 $q \overline{q} \rightarrow  q^{'} \overline{q}^{'}  $ (uds)  &   0.000e+00 & 0.000e+00 \\
 g g $\rightarrow c \overline{c}$        &   0.000e+00 & 0.000e+00 \\
 $q \overline{q} \rightarrow c \overline{c}$        &   0.000e+00 & 0.000e+00 \\
 g g $\rightarrow  b \overline{b}$        &   0.000e+00 & 0.000e+00 \\
 $q \overline{q} \rightarrow  b \overline{b}$        &   0.000e+00 & 0.000e+00 \\
\hline
 sum                                  &   9.336e+01 & 3.086e-02 \\
\hline 
\end{tabular} 
\end{center}
\begin{center}
{\it Table 4. Subprocesses and corresponding value of production cross section\\ in pp collisions at 13 TeV.}\\
\vspace*{3mm}
\begin{tabular}{|c|c|c|}
\hline 
  Subprocess                           &      sigma $\pm$ delta &   (estimated) (mb)   \\
 \hline
 non-diffractive                      &   5.642e+01 & 0.000e+00 \\
 A B $\rightarrow \text{ A B elastic           }$        &   2.226e+01 & 1.006e-07 \\
 A B $\rightarrow \text{ X B single diffractive}  $      &   6.416e+00 & 0.000e+00 \\
 A B $\rightarrow \text{ A X single diffractive}   $     &   6.416e+00 & 0.000e+00 \\
 A B $\rightarrow \text{ X X double diffractive}    $    &   8.798e+00 & 4.436e-08 \\
 g g $\rightarrow \text{ g g   }     $   &   4.580e-01 & 3.689e-02 \\
 g g $\rightarrow q \overline{q} (uds) $        &   1.338e-02 & 4.602e-03 \\
 $q g \rightarrow q g $        &   2.756e-01 & 3.181e-02 \\
 $q \overline{q}^{'} \rightarrow q \overline{q}^{'}$        &   2.437e-02 & 7.500e-03 \\
 $q \overline{q} \rightarrow \text{ g g         }$        &   0.000e+00 & 0.000e+00 \\
 $q \overline{q} \rightarrow q^{'}\overline{q}^{'}$ (uds)    &   0.000e+00 & 0.000e+00 \\
 g g $\rightarrow c \overline{c}$        &   5.188e-03 & 5.188e-03 \\
 $q \overline{q} \rightarrow c \overline{c}$        &   0.000e+00 & 0.000e+00 \\
 g g $\rightarrow b \overline{b}$               &   0.000e+00 & 0.000e+00 \\
 $q \overline{q} \rightarrow b \overline{b}$        &   0.000e+00 & 0.000e+00 \\
\hline
 sum                                  &   1.011e+02 & 4.977e-02 \\
\hline 
\end{tabular} 
\end{center}

Comparison of the data in Table 3 and Table 4 allows us to say that the pp interaction cross section increases with energy and there is a redistribution of the fraction of gluon-gluon and quark-antiquark processes with the formation of charm quark-antiquark and quark antiquark (uds) pairs, respectively at 13 TeV. In addition, we can conclude that  the largest part in the proton-proton collision is connected with non-diffractive processes than with elastic and double diffractive sub-processes.  The smallest fraction of the production cross section is accounted for by quark and gluon interactions.

\section{Conclusions}

We have considered proton-proton and proton-ion (Pb, Au) collisions at the energies of 5.02 TeV and 8.14 TeV. Using Pythia 8.3 program with Angantyr model for heavy ions we calculated the transverse momentum and  rapidity distributions  for K-meson and $\Lambda$-baryon at the energy of 5.02 TeV and 8.14 TeV. We received the following results:
\begin{itemize}
\item  There is the string influence on the p-ion interactions (production cross section larger for inclusion of strings);
\item  There is no string influence on the pp interactions;
\item  There is the impact parameter influence on p-Pb and p-Au interactions: for b=0.5 we have larger entries than for b=0 for K-meson as well as for Lambda baryon;
\item  There is no impact parameter influence on p-p interactions;
\item  There is the symmetric rapidity distribution for K-meson ejection;
\item  There is an asymmetry of the Lambda baryon ejection in the backward direction with an increase in the impact parameter;
\item  There is only small part of hard processes compared with non-diffractive processes for p-Au and p-Pb collisions;
\item  There is no part of hard processes in pp collisions compared with non-diffractive, elastic and diffractive processes for pp collisions at 8 TeV;
\item  There is some part of hard processes in pp collisions at 13 TeV;
\item  The sum of production cross sections in pp collisions is larger at 13 TeV.
\end{itemize}


\begin{thebibliography}{99}
\bibitem{1.}Glauber, R. J. Cross Sections in Deuterium at High Energies. Phys. Rev. V.100, Issue 1, 1955, p. 242-248.
\bibitem{2.}  Gribov V.N. Glauber Corrections and the Interaction between High-energy Hadrons and Nuclei. Sov. Phys. JETP 29 (1969) Issue 3, p. 483-487.
\bibitem{3.} Donnachie Sandy, Dosch Günter, Landshoff Peter and Nachtmann Otto. Pomeron Physics and QCD. Cambridge Monographs on Particle Physics, Nuclear Physics and Cosmology. Cambridge University Press, 2002, 347 p.
\bibitem{4.} Sj{\"o}strand, T. and Skands, P. Z. Transverse-momentum-ordered showers and interleaved multiple interactions. The European Physical Journal C, V.39, N 2, 2005, p.129-154.
\bibitem{5.}Bierlich, Christian
and Gustafson, G{\"o}sta
and L{\"o}nnblad, Leif
and Shah, Harsh. The Angantyr model for heavy-ion collisions in Pythia8. JETP, V.2018, N 10, 2018, 134 p. 
\bibitem{6.}ATLAS Collaboration. 
Measurements of long-range azimuthal anisotropies and associated Fourier coefficients for $pp$ collisions at $\sqrt{s}=5.02$ and 13 TeV and $p+\text{Pb}$ collisions at $\sqrt{{s}_{\mathrm{NN}}}=5.02$ TeV with the ATLAS detector. Phys. Rev. C, V. 96, Issue 2, 2017, 37 p.
\bibitem{7.}  ALICE Collaboration. Enhanced production of multi-strange hadrons in high-multiplicity proton--proton collisions. Nature Physics, V. 13, N 6, 2017, p. 535-539.
\bibitem{8.}CMS Collaboration. Search for a standard model-like Higgs boson in the mass range between 70 and 110 GeV in the diphoton final state in proton-proton collisions at s=8 and 13 TeV. Physics Letters B, V. 793, 2019, p. 320-347.
\bibitem{9.}B. Andersson and G. Gustafson and G. Ingelman and T. Sjöstrand.Parton fragmentation and string dynamics. Physics Reports, V. 97, N 2, 1983, p. 31-145.
\bibitem{10.}Andersson Bo,
 Gustafson G{\"o}sta
and Sj{\"o}strand Torbj{\"o}rn. A three-dimensional model for quark and gluon jets.
Zeitschrift f{\"u}r Physik C Particles and Fields, V. 6, N 3, 1980, p. 235-240.
\bibitem{11.}
Andersson, B.
and Gustafson, G.
and S{\"o}derberg, B. A general model for jet fragmentation. Zeitschrift f{\"u}r Physik C Particles and Fields, V. 20, N 4, 1983, p. 317-329.
\bibitem{12.} Christian Bierlich, G{\"o}sta Gustafson and Leif L{\"o}nnblad. Collectivity without plasma in hadronic collisions. Physics Letters B, V. 779, 2018, p. 58-63.
\bibitem{13.} Sj{\"o}strand Torbj{\"o}rn, Ask Stefan, Christiansen Jesper R., Corke Richard, Desai Nishita, Ilten Philip, Mrenna Stephen, Prestel Stefan, Rasmussen Christine O. and Skands Peter Z. An introduction to PYTHIA 8.2. Computer Physics Communications, V.
191, 2015, p. 159-177.
\bibitem{14.} L{\"o}nnblad Leif. The Angantyr model for heavy ions in Pythia8. The 28th International Conference on Ultra-relativistic Nucleus-Nucleus Collisions: Quark Matter 2019. Nuclear Physics A, V. 1005, 2021, p. 121873.

\end{thebibliography}
\end{document}